\documentclass[article]{aa}   

\usepackage{natbib}
\usepackage{graphicx}
\usepackage{txfonts}
\usepackage{array}
\usepackage{subfigure} 
\usepackage[skip=5pt]{caption} 

\begin{document}

\title{Multilayer modeling of porous grain surface chemistry}
\subtitle{I. The GRAINOBLE model}



\author{V. Taquet \and C. Ceccarelli \and C. Kahane}

\offprints{V. Taquet: vianney.taquet@obs.ujf-grenoble.fr}

\institute{UJF-Grenoble 1 / CNRS-INSU, Institut de Plan\'{e}tologie et d\textquoteright Astrophysique de Grenoble (IPAG) UMR 5274, Grenoble, F-38041, France}

\date{Received / Accepted}

\titlerunning{The GRAINOBLE model}
\authorrunning{V. Taquet et al.}

\abstract
{Mantles of iced water mixed with carbon monoxyde, formaldehyde, and
  methanol are formed during the so-called prestellar core phase. In
  addition, radicals are also thought to be formed on the grain
  surfaces, and to react to form complex organic molecules later on,
  during the so-called warm-up phase of the protostellar evolution. }
{We aim to study the formation of the grain mantles
  during the prestellar core phase and the abundance of formaldehyde,
   methanol, and radicals trapped in them. }
{We have developed a macrosopic statistic multilayer model that
  follows the formation of grain mantles with time and that includes
  two effects that may increase the number of radicals trapped in the
  mantles: i) during the mantle formation, only the surface
  layer is chemically active and not the entire bulk, and ii) the
  porous structure of grains allows the trapping reactive particles.  
  The model considers a network of H, O, and CO
  forming neutral species such as water, CO, formaldehyde, and
  methanol, plus several radicals. We ran a large grid of models to
  study the impact of the mantle multilayer nature and grain porous
  structure. In addition, we explored how the uncertainty 
  of other key parameters influences the mantle composition. }
{Our model predicts relatively high abundances of radicals, especially
 of HCO and CH$_3$O ($10^{-9}-10^{-7}$). In addition, the multilayer
 approach enables us to follow the chemical differentiation
 within the grain mantle, showing that the mantles are far from being
 uniform.  For example, methanol is mostly present in the outer
 layers of the mantles, whereas CO and other reactive species are
 trapped in the inner layers. The overall mantle composition depends
 on the density and age of the prestellar core as well as on some
 microscopic parameters, such as the diffusion energy and the
 hydrogenation reactions activation energy.  Comparison with
 observations allows us to constrain the value of the last two
 parameters (0.5-0.65 and $1500$ K, respectively) and provide 
 some indications on the physical conditions
 during the formation of the ices.}
{}
\maketitle

\section{Introduction}

Among the 150 molecules that have been detected in the interstellar
medium (ISM) so far, a significant number are complex organic
molecules (here-after COMs), carbon bearing molecules with more than five
atoms. Several COMs have been observed in large quantities for two
decades in the warm and dense hot cores of massive protostars
\citep{Blake1987}. They have received renewed interest in the last few
years after the detection of abundant COMs in solar-type protostars,
specifically in hot corinos \citep{Cazaux2003,Bottinelli2007}, and in
the clouds of the Galactic Center \citep{RequenaTorres2006}.

\begin{figure}[htp]
\centering
\includegraphics[width=88mm]{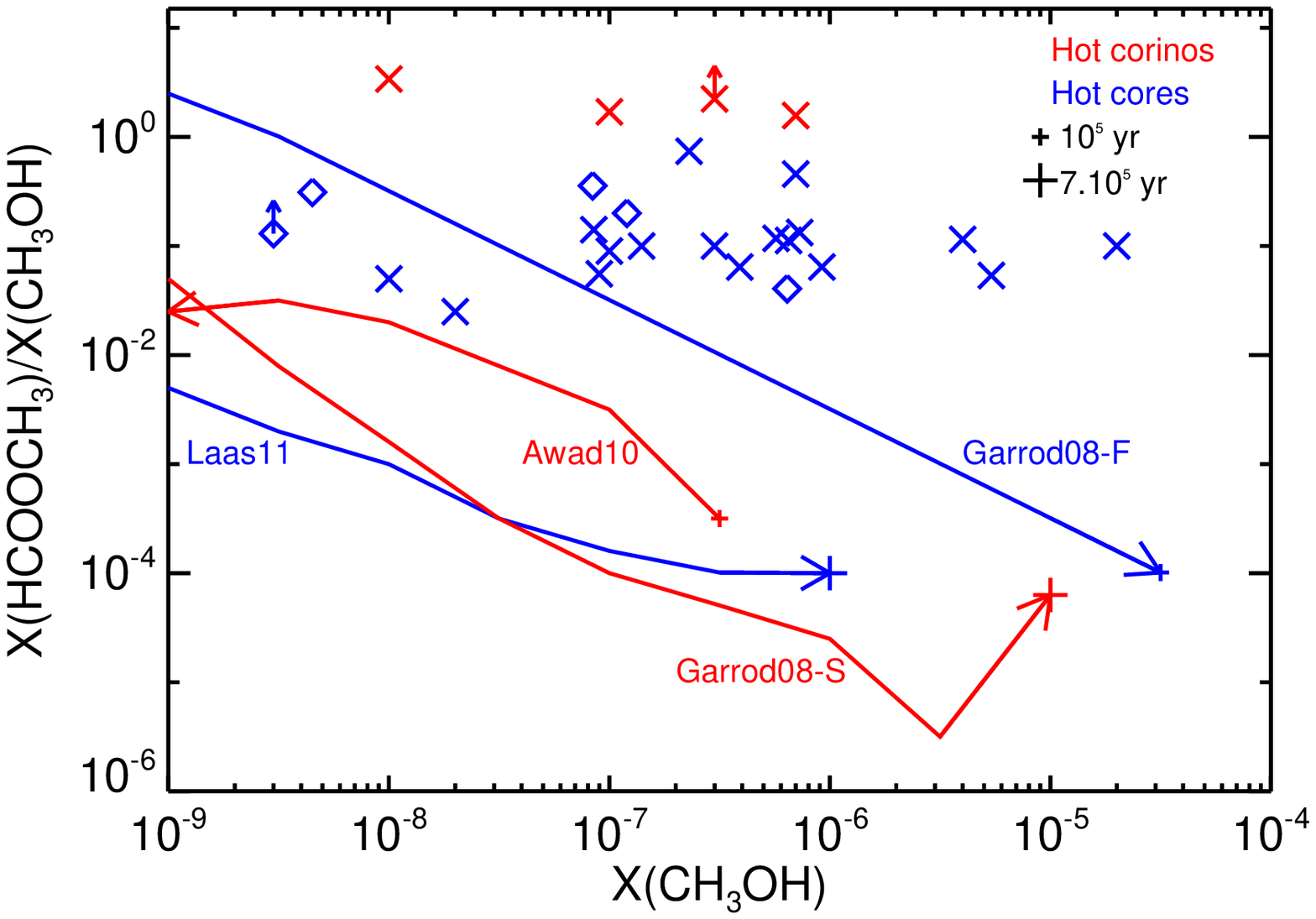}
\caption{Gas phase methyl formate to methanol abundance ratio as function of 
the gas phase abundance of methanol. Observations of hot corinos with single-dish telescopes
  are represented by red crosses: NGC1333-4A by
  \citet{Bottinelli2004}, IRAS16293 by \citet{Cazaux2003}, NGC1333-2A
  and -4B by \citet{Bottinelli2007}. Observations of hot cores with
  single-dish telescopes are represented by blue crosses: G34.3+0.15
  by \citet{Mehringer1996}, SgrB2(N) and SgrB2(M) by
  \citet{Nummelin2000}, G327.3-0.6 by \citet{Gibb2000}, OMC1 by
  \citet{Sutton1995}, G34.3+0.2, SgrB2(N), DR23(OH), W51, Orion
  Hot-Core by \citet{Ikeda2001}, AFGL2591, G24.78, G75.78, NGC6334,
  NGC7538, W3(H2O), W33A by \citet{Bisschop2007}. Observations of hot
  cores with interferometers are represented by blue diamonds:
  G34.3+0.15 by \citet{Macdonald1996}, G19.61-0.23 by \citet{Qin2010},
  Orion KL, G29.96 by \citet{Beuther2009}, and G47.47+0.05 by
  \citet{Remijan2004}. The red curves report the predictions of the
  methyl formate to methanol abundance ratio appropriate to the hot
  corinos case by Garrod et al. (2008, their Fig. 6) and Awad et
  al. (2010, their Fig. 5). The blue curves report the predictions
  appropriate to the hot cores case by Garrod et al. (2008, their
  Fig. 4) and Laas et al. (2011, their Fig. 5). The arrows represent
  the direction of the time in each model and the ticks refer to timescales.}
\label{methylformate_vs_methanol}
\end{figure}

Astrochemical models have shown that many COMs cannot be produced
efficiently in the gas phase. \citet{Horn2004}, for example, predicted
methyl formate abundances to be less than $10^{-10}$ by considering
formation pathways only in the gas phase, but this molecule has
been detected with abundances up to $10^{-6}$ (see references in the
caption of Fig. \ref{methylformate_vs_methanol}).  On the other hand,
several observational and experimental works have highlighted the
catalytic behaviour of interstellar grains for chemical reactions.
First, infrared (IR) observations of protostellar sources carried out with space
or ground-based telescopes have shown that formaldehyde and methanol
can be significant components of the grain mantles with fractional
abundances up to 30\% with respect to water in some cases
\citep{Gibb2004, Boogert2008}.  These mantles are believed to be
formed during the cold and dense prestellar core phase mainly via
hydrogenation and oxydation of CO and O. Second,
several laboratory experiments simulating cold cloud conditions have
confirmed the efficiency of the hydrogenation
processes. \citet{Watanabe2002} for instance have succeeded in
producing solid formaldehyde and methanol at low temperature 
($\sim$ $10$ K) via CO hydrogenation reactions in a CO-H$_2$O ice mixture.

Modelling the chemistry on the grain surfaces is therefore crucial if we
aim to understand the formation of COMs on the interstellar
grains. Over the past 30 years, several numerical methods based on
ab initio \citep{Allen1977}, Monte Carlo \citep{Tielens1982,
  Charnley1992}, or rate equations \citep{Hasegawa1992, Caselli1998a}
approaches have been developed. All these methods give a macroscopic
description of the problem, meaning that they follow the overall
behaviour of the mantle rather than the single particle.  This
method, based on the rate equations, though the least precise from a
physical point of view, is the fastest and allows studies of the
chemical evolution in objects with varying physical conditions as a
function of the time. In this context, time-dependent models have
been developed that predict the abundance of many COMs in hot
cores/corinos by considering two phases: a dense and cold pre-collapse
phase during which the grain mantles with simple hydrogenated species
are formed, and a warm-up phase, caused by the collapse, during which
heavier species, and particularly radicals, can react together on the
grain surfaces before sublimating into the gas phase
\citep{Garrod2006, Garrod2008a, Aikawa2008}. A key point of these
models is therefore the presence of radicals (OH, HCO, CH$_3$O)
trapped in the cold mantles. In these models, the radicals are assumed
to be mainly produced by the UV photodissociation of the neutral
species in the grain mantles.  \citet{Garrod2006} tested this
assumption by introducing reactions between OH, HCO, CH$_3$, CH$_3$O
to produce methyl formate, dimethyl ether, and formic
acid. \citet{Garrod2008a} then expanded the chemical network of
reactions between radicals to form other COMs such as ethanol,
glycolaldehyde, or acetic acid. Finally, \citet{Aikawa2008} and
\citet{Awad2010} have included the spatial distribution and evolution
of a collapsing cloud with the aim to estimate the size of typical
hot corinos such as the well-known source IRAS16293-2422.

Although the abundances of the simplest COMs, such as formaldehyde
(H$_2$CO) or methanol (CH$_3$OH), are now quite well predicted,
astrochemical models still fail to reproduce the abundance of more
complex ones, such as methyl formate (HCOOCH$_3$), for example.  This
molecule is assumed to be mainly formed on grains via the reaction
between HCO and CH$_3$O during the warm-up phase
\citep{Allen1977,Garrod2006}, a reaction in competition with the
hydrogenation reaction of CH$_3$O, which leads to methanol.  Figure
\ref{methylformate_vs_methanol} shows the observed and the predicted
methyl formate to methanol gas phase abundance ratios as function of
the gas phase methanol abundance. The plotted observations
  refer to the abundance values estimated in the articles cited in the
  figure caption, where the sizes are either directly estimated by
  interferometric observations or indirectly by other considerations
  for single-dish observations.  The plotted model predictions refer
  to gas phase abundances. The curves ``Garrod-F" and ``Garrod-S" 
  have been obtained from Fig. 4 and 6 of \citet{Garrod2008a} respectively,
  the curve ``Awad10" comes from Fig.  5 of \citet{Awad2010}, and the
  curve ``Laas11" from Fig. 5 of
  \citet{Laas2011}. Each point of the curves represents
  a different time or temperature (depending on the model). In these
  models, the methanol abundance increases with time, except in Awad
  et al.  (2010) who considered the destruction of the species in the
  gas phase. We remark that while the methanol abundances may suffer
  of the uncertainty on the source size and H$_2$ column density, the
  methanol to methyl formate abundance ratios are almost unaffected
  by these uncertainties. Therefore, unless the observed methanol
  abundance is always lower than 10$^{-7}$, which is certainly not the
  case, no model can reproduce the totality of
  observations. Specifically, the observed ratios are roughly
independent of the methanol abundance and are about 0.1 and 1 in hot
cores and hot corinos respectively, regardless of the telescope used
for the observations.  State-of-art astrochemical models consequently
underestimate the methyl formate to methanol abundance ratios and
always predict a decreasing ratio with increasing methanol abundance
(because methyl formate is in competition with methanol formation). 

Motivated by this unsatisfactory situation, we developed a new model,
called GRAINOBLE, whose ultimate goal is to simulate the
synthesis of COMs on the grain surfaces.  In this article, we focus on
the first phase, namely the formation of the mantle during the dense
and cold phase of the pre-collapse, the prestellar core phase. 
  Our model differs from most previous published models because it treats
  the multilayer and porous structure of the iced mantles.  In the
  mentioned previous models (Garrod \& Herbst 2006; Garrod et
  al. 2008; Aikawa et al. 2008; Awad et al. 2010; Laas et al. 2011),
  COMs are synthesized during the warm-up phase thanks to the
  increased mobility of radicals, formed on the grain surfaces during
  the previous cold phase. The abundance of radicals in the mantles
  during the cold phase is, therefore, a key point. One possibility,
  explored by the above models, is that the radicals are formed by the
  UV photodissociation of frozen species. However, since the impact of
  the UV photons on the surface chemistry is poorly understood (see
  the detailed discussion in \S 2.1), we explore here the possibility
  that radicals are synthesized even without photolysis and are trapped
  in the grain mantles, thanks to the multilayer mantle
  treatment.

The article is structured as follows. In the next section, we review
the models available in the literature, outlining their limits and pointing
out possible improvement. The concluding paragraph of the section
describes the improvements proposed by our new model GRAINOBLE. In
section \ref{sec:astrochemical-model} we describe the technical
details of GRAINOBLE. Section \ref{sec:results} reports the results of
the new computations, discussing a reference set of parameters and a
large grid of runs where several input parameters are varied. In
section \ref{sec:comp-with-prev} we compare the obtained results with
previous models with the twofold goal to validate our code and discuss
the different results from the different adopted physical
models and conditions. Section \ref{sec:comp-with-observ} compares the
GRAINOBLE predictions with the observations of ices and discusses the
constraints on the various parameters of the model.  Finally,
\S\ref{sec:concl-persp} concludes the article, emphasising the main
results and perspectives.

\section{Limits of the existing models and need for improvement}

\subsection{The physics of grain surface chemistry}

\noindent
\textit{a) Multilayer vers bulk chemistry} 

One of the main limits of many published models is that they do not
distinguish the chemical processes that occur in the mantle bulk and
on the surface. In these models, species that are buried in the mantle
can keep diffusing in the mantle and reacting with others even if
they are covered by new layers. Furthermore, particles landing on the
surface on top of the grain mantle can react with all particles,
regardless of their depth within the mantle.  Thus, heavy
reactive particles such as radicals can continue to react with
landing hydrogen atoms even though the radicals are buried deep in the
mantle bulk. The "bulk chemistry method" may therefore (grossly)
underestimate the abundance of radicals in grain mantles.

In contrast, the difference in the chemical behaviour of the mantle
and of the surface may lead to the formation of non-homogeneous grain
ices as suggested by the IR observations. \citet{Tielens1991}, for
instance, by observing two features in the CO bands,
suggested a polar mixture enriched by water in the bottom of the
mantle covered by a non-polar one mainly composed of CO located on the
outermost layers.  In addition, experiments have shown that UV and
cosmic rays irradiation can alter the chemistry of astrophysically
relevant ices composed of water, methanol, and other CO-bearing
molecules even in dark cold cloud conditions \citep{Gerakines1996,
  Bennett2007}. However, unlike highly energetic cosmic rays, which
can cross through the entire grain mantle, UV photons only penetrate
a limited number of monolayers, depending on the absorption cross-section
of molecules that constitute the ice \citep[see][for an illustrative discussion]{Gerakines2001}. 
Modelling the photolytic processes in ices requires,
therefore, a distinction between the surface and the bulk of the
mantle.

The desorption of grain mantles during the warm-up phase
depends on their chemical composition. By carrying out temperature
programmed desorption (TPD) experiments, \citet{Collings2004} demonstrated
that CO ices desorb at $\sim 20$ K, whereas methanol and water ices
desorb at temperatures higher than 100 K, for instance. Numerical
values of desorption energies were deduced from these experiments and
were incorporated in macroscopic rate equation models. Consequently,
the model of \citet{Garrod2008a} predicts a total desorption of the CO
reservoir at 25 K, whereas more refractory species are desorbed at
temperatures higher than 100 K . Actually, a significant amount of volatile 
ices can be trapped with highly polar molecules, such as water, in the mantle 
bulk. Thus, the desorption of such species is limited by the gradual evaporation of
the mantle main matrix, which depends on other molecules.

The continuous time random walk (CTRW) Monte-Carlo method, introduced by
\citet{Chang2005} and \citet{Cuppen2005}, strongly differs from the
macroscopic methods described above. This ``microscopic'' model
considers the actual hopping of a discrete number of particles from
one site to another but, unlike previous Monte Carlo methods, it takes
into account the actual position of particles so that the spatial
distribution is also included in the simulation. \citet{Cuppen2007}
and \citet{Cuppen2009} have explained the growth of ice monolayers in
dense clouds and showed a differentiation of the chemical composition
in the mantle. More particularly, \citet{Cuppen2009} showed that
under dense cloud conditions, CO can be trapped in the mantle by the
accretion of new species above it before reacting with other
particles. They also highlighted the survivability of very reactive
particles in the mantle bulk by this process. Unfortunately, this
method can only study simple chemical systems because of the CPU time
required by the Monte-Carlo algorithm, and a direct link with the gas
phase is not possible at present.

\citet{Hasegawa1993b} have attempted to treat separately the chemical
processes within the bulk and on the surface with a three-phase
model. They considered the formation of an inert mantle caused by the
accretion of particles onto it. They found that the differences
between the three-phase and the two-phase models are small for stable
species but become non-negligible for reactive species at long
($\geq 10^5$ yr) timescales. Furthermore, the values of desorption and
diffusion energies used in 1993 are lower than the values measured by
the experiments carried out during the past decade. Because higher
desorption and diffusion energies values tend to decrease the rate of
the chemical processes on grain surfaces, the impact of the
three-phase model on the frozen radical abundances would be
significantly higher with the new values.  However, at the time of the submission
of this article, this three-phase model has not been pursued in other works.
\\

\noindent
\textit{b) Porous grain surfaces} 

Astrochemical models used for predicting the COM formation have so far
only considered perfectly spherical and smooth grains.  However,
observations \citep{Mathis1996}, theoretical studies
\citep{Ossenkopf1993, Ormel2009}, and analyses of solar system bodies
have provided evidence for the fluffy and porous structure of
interstellar grains.  For example, \citet{Mathis1996} needed to use
interstellar grains with a substantial fraction of vacuum to fit the
observed extinction curves.  Numerical simulations of dust
coagulations carried out by \citet{Ossenkopf1993} from a MRN-like
grain size distribution have shown an increase of the size and the
porosity of the dust in molecular cloud conditions.  This fluffy
structure of interstellar grains may alter the surface chemistry
because grains can have micro-pores of small apertures compared to
their volume in which species can be trapped.
Thus, porous grains can increase the reactivity of chemical reactions
by trapping volatile particles such as atomic hydrogen in their pores.
\\

\noindent
\textit{c) Photolysis} 

As described in the introduction, the new generation of COMs
  formation models predicts that COMs are synthesized during the
  warm-up phase thanks to the increased mobility of radicals, formed
  on the grain surfaces during the previous cold phase \citep{Garrod2006,
  Garrod2008a,Aikawa2008,Awad2010,Laas2011}. The abundance of radicals in the
  mantles during the cold phase is, therefore, a key point. In these
  works, UV photons from cosmic rays play a major role in synthesising
  radicals during this phase. However, as for any model, a series of
  inevitable assumptions are adopted to treat the process.

  The first important assumption regards the photodissociation rates
  on the ices. They are assumed to be the same as those in the gas
  phase, computed by \citet{Sternberg1987, Gredel1989}. 
  However, there are various reasons to think that this
  may be a serious overestimate. First, grain surfaces could absorb
  part of the UV photon energy, ending up with a lower
  dissociation rate. Second, once the molecule is broken, if it is
  broken, being not in the gas phase but on a surface with almost no
  mobility, the photoproducts may recombine almost instantaneously. In addition, some
  products can even have enough energy to sublimate \citep[see for example][]{Andersson2006}. 
  Thus, the products of the
  photodissociation are likely different from those in the gas phase,
  and the overall formation rate of the assumed products (and
  radicals) may be severely lower than assumed. 

 A second point is related to the flux of CR induced UV photons in dense clouds.
 The exact value is fairly uncertain, 
 because it largely depends on the primary cosmic ray energy spectrum, and on the 
 grain extinction cross section, which are also uncertain \citep[see for example][]{Padovani2009}. For example, 
 \citet{Shen2004} have shown that the typical uncertainty on the low-energy
 cosmic ray spectrum leads to a variation of the UV flux of more than one 
 order of magnitude.

  Another important assumption regards the penetration of the UV
  photons through the bulk of the mantle. Unfortunately, there exists
  almost no observational or theoretical ground for that, and it is
  not obvious how realistic this assumption is. Indeed, the
  experimental work by \citet{Gerakines2000} suggests that the UV
  photons can only penetrate a limited number of layers, depending on
  the optical properties of the ice.

  Finally, even the branching ratios of species caused by the
  photodissociation in the gas phase are very poorly known \citep[e.g,][]{Laas2011}. 
  In this context, various authors omit the photolysis
  in their model \citep[e.g,][]{Chang2007,Cuppen2009}.

\subsection{The computational methods}\label{sec:comp-meth}

The rate equations method allows us to study the evolution of the
chemical composition of grain mantles via one temporal differential
equation for each species. This approach allows us the use of complex
chemical networks that include photolytic processes involving hundreds
of species, and allows us to directly link the gas phase and the grain
surfaces. However, rate equations are based on the evolution of
densities of species in gas-phase and on grain surfaces, which are
average values. Therefore this approach can be inaccurate when the 
number of particles on the grain becomes low because of the finite surface 
of grains. \citet{Caselli1998a} took into account the discrete nature
of the individual composition of grain mantles by modifiying the
reaction rate coefficients, while \citet{Garrod2008a} went further by
modifying the functional form of the reaction rates.
\citet{Biham2001} and \citet{Green2001} introduced a stochastic method
based on the resolution of the master equations (ME), in which the system
solves the time derivatives of the probability of each discrete mantle
composition. In spite of its good accuracy for grain surfaces, this
method can only be used for small chemical networks since the number
of equations grows exponentially with the number of
species. \citet{Lipshtat2003} have thus introduced another method
based on the time derivatives of the moments to study the formation of
molecular hydrogen, while \citet{Barzel2007} have extended this method
for more complex chemical networks. The comparisons between the
methods introduced here show that the rate equations method tends to
overestimate the reaction rates and the abundance of the reactants, 
especially when they are in low quantities.

\subsection{Need for improvement: the GRAINOBLE model}

Although the past few years have seen a huge improvement in the
modelling of grain surface chemistry, models are not yet able to
accurately reproduce observations
(e.g. Fig. \ref{methylformate_vs_methanol}). Therefore, something is
still missing in those models. In the following, we review what are,
in our opinion, the areas where improvements are needed and
possible. Our new grain surface model, GRAINOBLE, has been developed
to include the improvements described below. 
\\

\noindent
{\it a) Multilayer structure of the ices}

The first obvious limitation of the Herbst, Garrod, and collaborators class of
models, the present state of the art, is that they do not take into
account the multilayering formation of the ices, which leads to the
differentiation observed in the interstellar ices
\citep{Tielens1991,Pontoppidan2003} that is also predicted by the microscopic
Monte-Carlo (MC) models of Cuppen and collaborators, and to the potential
survival of reactive particles in the mantle bulk. On the other
hand, the microscopic MC treatment is too cumbersome in computer time
and difficult, if not impossible, to apply to model realistic
cases. GRAINOBLE uses the rate equation method developed by Herbst and
collaborators (which is at the base of the Garrod models) in a
way that permits us to follow the multilayering structure of the
ice. Therefore, it benefits from the advantages of the rate
equation method
(computer speed) and of the microscopic MC (multilayer) approach. \\

\noindent
{\it b) Porosity of the grains}

Laboratory experiments, numerical simulations of grain coagulation,
mantle formation, and astronomical observations all demonstrate that
interstellar grains are porous \citep{Jones2011}. As previously
discussed, the presence of pores creates traps for the molecules and
atoms on the grain surfaces, leading to an enhancement of reactivity
on the grains. GRAINOBLE includes pores, following the treatment of
\cite{Perets2006}. The number and area of the pores are treated
as parameters (section \ref{sec:astrochemical-model}).\\

\noindent
{\it c) Multiparameter approach}

Many parameters in the microphysics of the problem have highly
uncertain values, like the various activation barriers of key chemical
reactions, or the barriers against diffusion of adsorbed
particules. Additionally, several macroscopic parameters are also
either uncertain or vary depending on the astronomical object. In
particular, it is worth mentioning that the densities and the
temperatures of the prestellar cores depend on their age and their
surroundings.

\section{The GRAINOBLE model}\label{sec:astrochemical-model}

\subsection{General description}

Following the work of \citet{Hasegawa1992}, four main processes
occuring on grains are taken into account.
\begin{itemize}
\item [1)] Gas-phase particles can accrete onto the grains, 
  which are considered spherical. For a given gas-phase species $i$, the
  accretion rate is a function of the thermal velocity of the
  gas-phase species $v(i)$, of the cross section of the grains $\sigma(a_d)$, of the density of 
  the grains $n_d(a_d)$ where $a_d$ is the grain diameter, and of the
  sticking coefficient $S(i)$. We assumed the sticking coefficient
  estimated by \citet{Tielens2005} for atomic hydrogen and a sticking
  coefficient equal to 1 for heavier particles, based on the TPD
  experimental work of \citet{Bisschop2006}. For dark cloud
  conditions, particles constituting the ice are bound mainly via the
  physisorbed van de Waals interactions.
\item [2)] Once the particles are stuck onto the grains, they can
  diffuse along the surface via thermal hopping according to the
  Boltzmann law, which gives the hopping rate $R_{hop}$. Following
  experimental work of \citet{Katz1999}, we neglected tunelling diffusion.
\item [3)] Physisorbed particles can react only via the
  Langmuir-Hinshelwood \citep{Hinshelwood1940} process, in which two
  species can react when they meet in the same site. The reaction rate
  $R_{r}$ is given by the product between the diffusion rate (i.e, the number 
  of times per second that a species sweeps over a number of sites equal to
  the number of sites of the layer) and the
  probability of the chemical reaction, which is a function of its
  activation energy.
\item [4)] Surface species can desorb only via thermal processes,
  \textit{i}) the ``classic" thermal process caused by the thermal balance
  of the grain, \textit{ii}) the cosmic ray induced desorption
  process in which cosmic rays heat the grains, as suggested by
  \citet{Leger1985}, whose heating rate was computed by
  \citet{Hasegawa1993a}. The sum of these two thermal desorption rates gives 
  the total evaporation rate $R_{ev}$. 
\end{itemize}
A full list of the symbols used in this work is provided in the appendix.

\subsection{Porosity}\label{sec:porosity}

To model the impact of the porous structures of interstellar grains,
we introduce two types of sites: the \textit{non-porous}, and the \textit{porous}
sites, following \cite{Perets2006}.
\begin{figure}[htp]
\centering
\includegraphics[width=88mm]{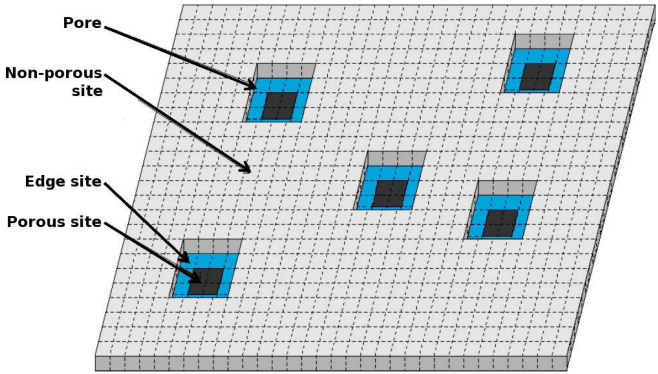}
\caption{Schematic view of a portion of the mantle layer as modelled by
  GRAINOBLE.  All pores are assumed to be square, to have the same
  size, and no walls (see text). The edges sites, that connect the
  non-porous surface and the pores are depicted in blue. In this example, each
  pore is constituted by 16 sites, of which the 12 blue sites are edge sites, and the
  fraction of area occupied by the pores is 0.1.}
\label{schema_pores}
\end{figure}
The {\it non-porous} sites form a smooth surface that is perfectly regular,
where particles interact directly with the gas phase. The gas phase
species can accrete onto the grain, and the adsorbates can desorb into
the gas phase.
The {\it porous sites} correspond to sites where no direct
accretion and desorption are possible. They are filled up only by
diffusion from non-porous sites at the edge of the pores.
The porous sites cover a fraction $F_{por}$ of the grain surface. For
the sake of simplicity, all the pores are assumed to be square and to
have the same size, constituted by $N_{pore}$ sites, where $N_{pore}
\geq 4$.  The fraction of the grain surface occupied by the pores,
$F_{por}$, is poorly known because, to date, no experimental
measurements or numerical estimates have been attempted. Therefore 
we ran models with four values for $F_{por}$: 0, 0.3, 0.6, and 0.9, 
to cover a large range of possibilities, 

The fraction of the edge porous sites, $F_{ed}$, on the grain depends
then on $F_{por}$  and $N_{pore}$ as follows:
\begin{equation} 
F_{ed} = 4 \cdot F_{por} \frac{\sqrt{N_{pore}} - 1}{N_{pore}} .
\end{equation}
A posteriori, the actual value of $N_{pore}$ does not influence the results 
significantly because what matters is mostly the fraction
of area occupied by the pores and not the size or the number of
pores. In the following, we assumed $N_{pore}=9$.

The approach that we used for modelling the grain porosity is sketched
in Fig.  \ref{schema_pores}.

\subsection{Multilayer approach}\label{sec:multilayer}

\begin{figure}[htp]
\centering
\includegraphics[width=88mm]{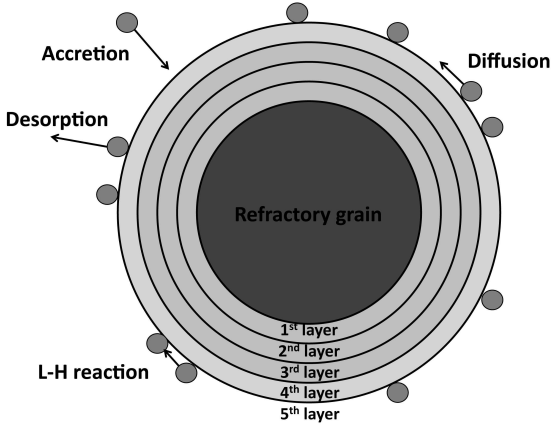}
\caption{Schematic view of the multilayer approach adopted by the
  GRAINOBLE code. The inner layers of the mantle are chemically inert,
  whereas the layer at the surface is chemically reactive (see text).}
\label{schema_layer}
\end{figure}

In order to distinguish the chemistry on the surface and in the bulk
of the mantle, we used a multilayer approach. In practice, the code
follows the chemical processes on a single layer, the one at the
surface of the mantle.  In this layer, particles can accrete, diffuse,
react with each other and desorb, depending on whether they are
  on non-porous or porous sites. The mantle layer growth is treated at
  the same time as the chemistry. Specifically, at each time $t$, the
  chemistry is followed by solving the set of differential equations
  described in \S 3.9 between $t$ and $t+ \Delta t$ and a new number
  of particles on the grain is computed at $t+ \Delta t$. The layer is
  considered chemically inert as soon as the number of particles that
  are on the layer is equal to the number $N_s(a_d)$ of (porous plus
  not-porous) sites of the layer, where $N_s(a_d)$ is given by the
ratio between the grain surface area and the site area $d_s^2$ (see \S
\ref{sec:part-surf-surf}). Because the porous sites are populated
  only via diffusion (\S 3.2), they fill up more slowly than the
  non-porous sites. It is therefore possible that the layer is considered
  inert when some pores are still empty. In this case, the number of
  non-porous sites is “artificially” increased. Although this is
  somewhat arbitrary, a posteriori this assumption has almost no
  impact on the results, as described in \S 4.3. When the layer
  becomes inert, the code memorizes its composition and a new and
reactive layer is started. Note that no chemical exchange is
  allowed between layers. Finally, the code also takes into account
the growth of the grain size and of the number of sites of the new
layer by assuming that the thickness of the layer is equal to the site
size $d_s$. All layers have, therefore, the same thickness, equal
  to the site size. Figure \ref{schema_layer} gives a schematic view
of this approach.

\subsection{Chemical network}\label{sec:chemical-network}

We adopted the network based on the work by \citet{Allen1977} and
\citet{Tielens1982}, updated for the new theoretical and experimental
results and adapted to model the formation of the main constituents of
intestellar ices.  As previously mentioned in the introduction,
observations of ices have shown that grain mantles are mainly composed
of O- and CO- bearing molecules such as water, CO, CO$_2$, or
methanol. Molecules such as NH$_3$ and CH$_4$ produced by the
hydrogenation of N and C can also be found, but only in smaller
quantities \citep{Gibb2004}. In this work, we therefore neglected the
formation of ammonia and methane.  In addition, for the
  reasons explained in \S 2.1, we also do not consider photolytic
  processes. Our model, therefore, considers the formation of stable
  species (e.g, formaldehyde and methanol) and radicals (e.g,
  HCO, and CH$_3$O) via hydrogenation of frozen CO. Note that we did
  not consider CH$_2$OH because, in absence of photolytic processes, it is
  formed via the hydrogenation of H$_2$CO. Theoretical studies have
  shown that the reaction H$_2$CO + H $\rightarrow$ CH$_2$OH has an
  activation barrier (10.5 kcal/mol) more than twice the activation
  barrier leading to CH$_3$O (4--6 kcal/mol) \citep{Sosa1986, Walch1993, Woon2002}. 
  We therefore neglected CH$_2$OH in the present study.
For the H$_2$O formation, we adopted the new reaction scheme  based on
the experimental works by \citet{Dulieu2010}, \citet{Ioppolo2010} and
\citet{Cuppen2010}, who estimated the energy barriers of several
reactions.
The recombination reaction of hydrogen atoms, forming molecular hydrogen, is also 
taken into account. Since H$_2$ has a low binding energy with respect to water ice 
\citep{Tielens1987, Perets2005}, and since its reaction of formation releases
 chemical energy \citep{Hollenbach1970}, we assumed
that the H$_2$ molecules are desorbed in the gas phase as soon as they are formed.
Finally, IR observations have shown that
CO$_2$ may be a major component of ices with abundances reaching 30 \%
with respect to to water in some cases \citep{Dartois1999, Pontoppidan2008}. However, the
formation pathways of carbon dioxide, either by a direct surface route
or by UV photolysis, in such cold conditions are still not well
understood. Accordingly, we decided to neglect the formation of this
molecule. 

Table \ref{table_energies} lists the chemical species considered in
our network with their binding energies relative to an
amorphous water ice substrate. Indeed, interstellar grains located in
prestellar-cores are already covered by a polar ice mainly composed of
water, formed during an earlier phase when atomic abundances are still
high, before the depletion of CO.
The network used for the grain surface chemistry is reported in Table
\ref{chemical_network}.

\begin{table}[htp]
\centering
\caption{List of species and binding energies assuming that the ice
  matrix is mainly water ice.}
\begin{tabular}{c c c}
\hline
\hline
Species & $E_b$(water ice) (K)  \\
\hline
H & 450 \tablefootmark{a} \\
O & 800 \tablefootmark{b} \\
CO & 1150 \tablefootmark{c} \\
OH & 2820 \tablefootmark{d} \\
H$_2$O & 5640 \tablefootmark{e} \\
O$_2$ & 1000 \tablefootmark{c} \\
O$_3$ & 1800 \tablefootmark{f} \\
HO$_2$ & 5570 \tablefootmark{c}  \\
H$_2$O$_2$ & 6300 \tablefootmark{f} \\
HCO & 1600 \tablefootmark{f} \\
H$_2$CO & 2050  \tablefootmark{f} \\
CH$_3$O & 5080 \tablefootmark{f} \\
CH$_3$OH & 5530 \tablefootmark{c} \\
\hline
\end{tabular}
\label{table_energies}
\tablebib{$^{(a)}$ \citet{Hollenbach1970}; $^{(b)}$ \citet{Tielens1982}; $^{(c)}$ \citet{Collings2004}; $^{(d)}$ \citet{Avgul1970}; $^{(e)}$ \citet{Speedy1996}; $^{(f)}$ \citet{Garrod2006}}
\end{table}

\begin{table}[htp]
\centering
\caption{Surface reactions and activation energies.}
\begin{tabular}{c l c}
\hline
\hline
Number & Reaction & $E_a$ (K) \\
\hline
1 & O + O $\longrightarrow$ O$_2$  & 0 \\
2 & O + O$_2$ $\longrightarrow$  O$_3$  & 0 \\
3 & H + H $\longrightarrow$ H$_{2,gas}$  & 0 \\
4 & H + O $\longrightarrow$ OH  & 0 \\
5 & H + OH $\longrightarrow$ H$_2$O   & 0 \\
6 & H + O$_2$  $\longrightarrow$ HO$_2$  & 0 \\
7 & H + HO$_2$ $\longrightarrow$ H$_2$O$_2$   & 0 (36\%)\tablefootmark{a}\\
8 & H + HO$_2$ $\longrightarrow$ OH  + OH   & 0 (57\%)\tablefootmark{a} \\
9 & H + H$_2$O$_2$ $\longrightarrow$ H$_2$O + OH  & 1400\tablefootmark{b} \\
10 & OH + OH $\longrightarrow$ H$_2$O$_2$ & 0 \\
11 & H + O$_3$ $\longrightarrow$ O$_2$ + OH & 0 \\
12 & H + CO $\longrightarrow$ HCO & $400$\tablefootmark{c} $< E_a < 2500$\tablefootmark{d} \\ 
13 & H + HCO $\longrightarrow$ H$_2$CO & 0 \\
14 & H + H$_2$CO $\longrightarrow$ CH$_3$O & $400$\tablefootmark{c} $< E_a < 2500$\tablefootmark{d}\\
15 & H + CH$_3$O $\longrightarrow$ CH$_3$OH & 0 \\
\hline
\end{tabular}
\label{chemical_network}
\tablebib{$^{(a)}$ branching ratios measured by \citet{Cuppen2010}; $^{(b)}$ \citet{Klemm1975}; $^{(c)}$ \citet{Fuchs2009}; $^{(d)}$ \citet{Woon2002}}
\end{table}

\subsection{Gas phase initial abundances} \label{sec:gasphase}

The observed and predicted gas-phase abundance of atomic carbon is much lower
than CO or atomic oxygen abundances. Therefore, we
considered only the accretion of gaseous H, O, and CO onto the grain
mantles.

In dark cloud conditions, the abundance of atomic hydrogen in the gas phase results from a
balance between the cosmic rays ionization of H$_2$ on the one hand
and the accretion on dust grains (which react to form H$_2$ or heavier
iced species) on the other hand. At steady state, the density of H is
therefore given by the ratio between these two processes
\citep{Tielens2005}, as follows:
\begin{equation} 
  n(H) = \frac{2.3 \cdot \zeta_{CR} \cdot n(H_2)}{2\cdot v(H) \cdot \sigma_d \cdot X_d \cdot n(H_2)} \label{eq_nH}
\end{equation}
where $v(H)$ is  thermal velocity of atomic hydrogen. The resulting H density is independent of the density and equal
to 1.2 cm$^{-3}$ at 10 K, for a cosmic rays ionization rate $\zeta$ of
$3 \cdot 10^{-17}$ \citep{Caselli1998b}, a grain abundance $X_d$ of $1.33
\cdot 10^{-12}$ relative to H nuclei and for a grain radius $r_d$ of 0.1 
$\mu$m ($\sigma_d = \pi \cdot r_d^2$), giving 
$n_{d} \cdot \sigma_{d}/n_H = 4.2 \cdot 10^{-22}$ cm$^{2}$.  Observations
and astrochemical models show that CO is a very stable species and the
most abundant molecule after H$_2$ in dark clouds. Following the
estimates of \citet{Frerking1982}, we assumed an initial abundance
of CO of $4.75 \cdot 10^{-5}$ with respect to H nuclei. This abundance
decreases with time because of the freeze-out onto the grain mantles.
Finally, the abundance of atomic oxygen in the gas phase remains
poorly constrained. Observations of the [OI] 63 $\mu$m line obtained
by \citet{Caux1999, Vastel2000, Lis2001} with the Infrared Space
Observatory (ISO) have measured abundances higher than $10^{-4}$,
suggesting that all the gaseous oxygen is in atomic form. 
But astrochemical models predict an atomic oxygen abundances of
$6 \cdot 10^{-5}$ or less. The initial abundance of O is therefore
considered as a free parameter and can vary between $2 \cdot 10^{-4}$
and $2 \cdot 10^{-5}$.

\subsection{Physical conditions}\label{sec:physical-conditions}

As supported by observational and theoretical arguments, the physical
properties of prestellar cores vary substantially with the age and the
mass of the object \citep[see][]{diFrancesco2007, Bergin2007}. The
 densities $n_H$ evolve typically from $\sim 10^4$ to $\sim 10^6$
cm$^{-3}$, while the gas kinetic temperature (equal to the grain
temperature in most cases) varies between 7 and 20 K,
depending on the type of the object. Indeed, massive stars are
suspected to be formed from warmer prestellar cores than low mass
ones. Consequently, we included several values of the density and the temperature, listed
in Table \ref{table_grid}, in our parameter grid.

Additionally, as shown by numerical simulations \citep{Ossenkopf1993,
  Ormel2009} and observations \citep{Stepnik2003, Steinacker2010}, the
coagulation of grains occuring in molecular clouds and in prestellar
cores tend to increase their size. That is why the grain size distribution,
which is assumed to follow a broad nearly power-law in diffuse interstellar
clouds \citep{Mathis1977}, tends to evolve with time, the bigger
grains dominating most of the dust mass. We assumed all
interstellar grains to have the same
diameter $a_d$, and we considered three values: 0.1, 0.2, and 0.3 $\mu$m respectively.
The grain abundance $X_d$ changes in accordance with the grain size, 
to keep a dust-to-gas mass ratio of 1 \%.

\subsection{Other key parameters of GRAINOBLE}\label{sec:part-surf-surf}

In addition to the above described parameters, astrochemical models of
grain chemistry are based on several other parameters related to the
particle-surface, surface particle-particle interaction and surface
properties.  We list below three additional key parameters of the
GRAINOBLE model whose values are poorly constrained
because of the difficulties involved in measuring or computing them.
We considered them as free parameters in a
range defined by the different values found in the literature. \\

\noindent
{\it a)} Diffusion energy to binding energy ratio $E_d/E_b$

Theoretical calculations of barriers against diffusion for physical
adsorption $E_d$ on perfectly smooth surfaces have been carried out by
\citet{Jaycock1986} and have shown that $E_d$ is typically about 30 \%
of the binding energy $E_b$.  However, ``real" surfaces, which show
defects, irregularities, or steps, tend to increase this ratio
\citep{Ehrlich1966}. Accordingly, several experimental studies have been
carried out to constrain the diffusion barrier on
astrophysically relevant surfaces.  By fitting experimental formation
of molecular hydrogen on refractory bare surfaces and on amorphous
water ices with a rate equation model, \citet{Katz1999} and
\citet{Perets2005} experimentally found high $E_d/E_b$ ratios (0.77
and $\sim 0.85$ respectively). However, \citet{Collings2003},
\citet{Ulbricht2002}, and \citet{Matar2008} estimated lower energy
ratios (of about 0.5) for CO and D on water ice, and for atomic oxygen
on carbon nanotubes. Clearly, the energy $E_d/E_b$ ratio is
highly uncertain and strongly depends on the composition and the
structure of the substrate. Three values of the $E_d/E_b$ energy ratio
have been considered here: 0.5, 0.65, and 0.8.\\

\noindent
{\it b)} Activation energy $E_a$

The activation energy values of the CO and H$_2$CO hydrogenation
reactions are also very uncertain. Theoretical studies of these
reactions on an icy mantles via calculations of quantum chemistry
carried out by \citet{Woon2002} and \citet{Goumans2011}, showed that
the activation energies are about 2000-2500 K, and depend slightly on the
presence of water molecules around the reactants. In contrast to these 
theoretical values, experiments show that these reactions take place
at low temperature \citep{Watanabe2002}. \citet{Fuchs2009}, by
modelling the experiments of a hydrogen deposition on a CO ice with a
CTRW method, found values of about 400-500 K for both reactions. We
therefore consider these barriers as a free parameter. Three values of
activation energies, assumed to be equal for the H+CO and H+H$_2$CO
reactions, have been considered: $E_a = 400, 1450, 2500$ K.\\

\noindent
{\it c)} Site size $d_s$

The surface density of sites $s$, i.e. the number of sites per surface
unit given in sites x cm$^{-2}$, depends on the structure of the
surface and its composition. Most astrochemical models assume a
value of $s = 10^{15}$ cm$^{-2}$ which has been measured
experimentally by \citet{Jenniskens1995} for a high-density amorphous
water ice. By assuming that the distance between two sites $d_s$ is
constant along the surface, $d_s$ can be easily deduced from the
surface density $s$ ($d_s = 1/\sqrt{s}$), and is equal to $3.2$
$\AA$. However, \cite{Biham2001} found a value of $s = 5 \cdot
10^{13}$ cm$^{-2}$ ($d_s = 14$ $\AA$) for an amorphous carbon surface
and \citet{Perets2006} assumed a value of $s = 5 \cdot 10^{15}$
cm$^{-3}$ ($d_s = 1.4 $ $\AA$) in their model of porous grains. We
therefore treat $d_s$ as a free parameter, with values equal to 1.4,
4.2 and 7 $\AA$. \\

\subsection{Multi-parameter approach} \label{sec:multiparam}

To summarize, the GRAINOBLE model depends on eight key parameters
presented in the previous paragraphs and listed in
Tab. \ref{table_grid} along with the range and values considered in
this study for each of them.  The values reported in bold are used in
the ``reference" model (\S \ref{sec:results}). They correspond to the
average physical conditions of prestellar cores (density, temperature,
and grain size), to the average values found in the litterature (site size, 
diffusion energy to binding energy ratio, initial abundance of atomic oxygen), 
to the values of parameters used  by most of previous  astrochemical models 
(porosity factor) or to the most recent values 
measured by experiments (activation energy).

In total, we ran a grid of 17496 models varying these eight free parameters.

\begin{table}[htp]
\centering
\caption{List of the key free parameters of GRAINOBLE and the value range.}
\begin{tabular}{c c}
\hline
\hline
Parameter & Values  \\
\hline
Density $n_H$ & $10^{4}$ - $\mathbf{10^5}$ - $10^{6}$ cm$^{-3}$ \\
Temperature $T_g = T_d$ & 10 - \textbf{15} - 20 K \\
Initial oxygen abundance $X$(O)$_{ini}$ & $2 \cdot 10^{-5}$ - $\mathbf{6 \cdot 10^{-5}}$ - $2 \cdot 10^{-4}$ \\
Grain size $a_{d}$ & 0.1 - \textbf{0.2} - 0.3 $\mu$m \\
Energy ratio $E_d/E_b$ & 0.5 - \textbf{0.65} - 0.8 \\
Porosity factor $F_{por}$ & \textbf{0} - 0.3 - 0.6 - 0.9 \\
Site size $d_s$ & 1.4 - \textbf{4.2} - 7 $\AA$ \\
Activation energy $E_a$ & \textbf{400} - 1450 - 2500 K \\
\hline
\end{tabular}
\tablefoot{Bold values mark the values adopted in the reference
  model (see text, \S \ref{sec:results}).}
\label{table_grid}
\end{table}

\subsection{Computational aspects} \label{sec:equadiff}

The code solves three sets of differential equations: the first one
describes the evolution of the densities of species (other than H, whose density is
given by eq. \ref{eq_nH}) in the gas phase,
the other two describe the chemical composition of the mantle
outermost layer on the non-porous surface and within the pores.  The
equations giving the evolution of the chemical composition of the
mantle layer do not depend on the grain size and are expressed in
monolayers/sec (MLs/s) where a monolayer is given by the number of particles
of the considered species divided by the number of sites of a layer.
The equations are the following:

\begin{eqnarray}
 \frac{dn_{g}(i)}{dt} = &-& S(i) \cdot v(i) \cdot n_{g}(i)  \cdot \sigma(a_{d}) \cdot n_{d}(a_{d})   \nonumber \\
   & + & R_{ev}(i)  \cdot P_{np}(i) \cdot N_{s}(a_d) \cdot n_d(a_d)  
   \label{ed_gas}
\end{eqnarray}
\begin{eqnarray}
\frac{dP_{np}(i)}{dt} = & & \frac{1}{4} \cdot S(i) \cdot d_{s}^2 \cdot v(i) \cdot n_g(i) - R_{ev}(i) \cdot P_{np}(i)  \nonumber \\
& - & \frac{F_{ed}}{F_{np}} \cdot R_{hop}(i) \cdot P_{np}(i) + \frac{F_{ed}}{F_{por}} \cdot R_{hop}(i) \cdot P_{por}(i) \nonumber \\
& + &  \sum_{i_f} R_{r,ext}(i_f) \cdot P_{np}(i_{r1}) \cdot P_{np}(i_{r2}) \nonumber \\
& - &  \sum_{i_d} R_{r,ext}(i_d) \cdot P_{np}(i_{r1}) \cdot P_{np}(i_{r2}) 
   \label{ed_ext}
\end{eqnarray}
\begin{eqnarray}
\frac{dP_{por}(i)}{dt} = &+& \frac{F_{ed}}{F_{np}} \cdot R_{hop}(i) \cdot P_{np}(i) - \frac{F_{ed}}{F_{por}} \cdot R_{hop}(i) \cdot P_{por}(i) \nonumber \\
& + &  \sum_{i_f} R_{r,in}(i_f) \cdot P_{por}(i_{r1}) \cdot P_{por}(i_{r2}) \nonumber \\
& - &  \sum_{i_d} R_{r,in}(i_d) \cdot P_{por}(i_{r1}) \cdot P_{por}(i_{r2}) 
   \label{ed_in}
\end{eqnarray}

where $P_{np}(i)$ and $P_{por}(i)$ are the surface populations of the
species $i$ (in MLs) on the non-porous surface and within the pores
respectively, and $n_{g}(i)$ its gas-phase density.  The first and
second terms in equation \ref{ed_gas} describe the accretion rate and
the evaporation rate.  The first and second
terms of equation \ref{ed_ext} describe the accretion and the
evaporation rates. The third and fourth terms describe the exchange
rate between the non-porous surface and the pores via diffusion.
The fifth and sixth terms describe the production and
the destruction rates of $i$ via reactions on the non-porous
surface. 
Similarly, the first and second terms of equation \ref{ed_in} describe 
the exchange rate between the non-porous surface and the pores via diffusion.
The third and fourth terms describe the production and
the destruction rates of $i$ via reactions in the pores.
Note that the evolution of composition within all the pores can be treated with a
single equation because all pores are assumed to have the same
size. The equations giving the evolution of each pore can be deduced
from equation \ref{ed_in} by multiplying it with a simple factor of
proportionality.

\subsection{What the model does not consider}

We aim to study the impact of the multilayer behaviour and
the porous structure of the grains on the mantle formation and
 its chemical composition. To this end, we considered thermal processes
only.  Therefore we did not consider the
desorption processes caused by exothermic surface reactions
\citep{Garrod2007}, and the photolytic desorption
\citep{Oberg2009a, Oberg2009b, Arasa2010} in spite of their relative 
possible importance. Neither did we account for the
photodissociation processes on the ice in
this version of GRAINOBLE owing to the reasons given in Sect. 2.1.

\section{Results}\label{sec:results}

We ran a grid of 17496 models, varying the eight
free parameters listed in Table \ref{table_grid}. In this section, we
describe the results of our model. We start by describing the results
relative to the reference model, and then we describe how the results
change when each of the parameters changes. The analysis is based on
the distribution of the obtained mantle abundances when all
parameters are varied, except for the parameter under consideration. This
allows us to understand if and by how much the considered parameter
influences the results.  The plot of the multilayer and bulk distributions
 is shown in section \ref{sec:multilayer-vs-bulk},
while the distributions of other parameters (using the multilayer approach only)
are shown in the appendix. 
Finally, Table \ref{summary_grid} summarises the main effects on the
resulting mantle abundances caused by each parameter.

\subsection{The reference model}\label{sec:reference-model}

\begin{figure*}[htp]
\centering
\includegraphics[width=180mm]{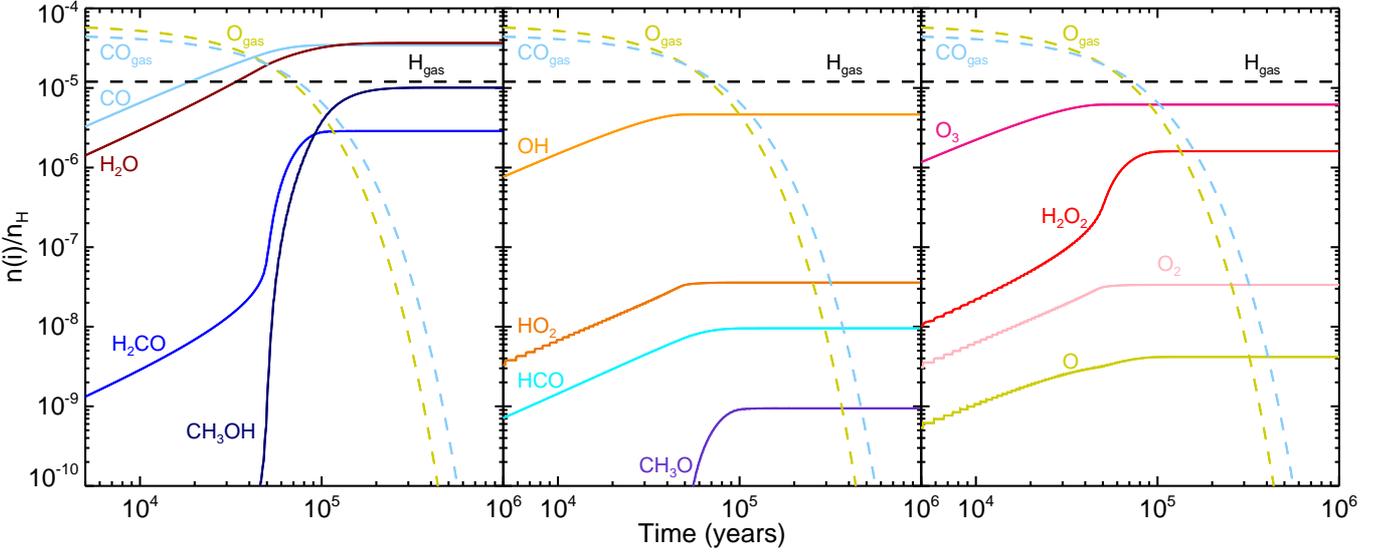}
\caption{Abundance of species in interstellar grain mantles (solid)
 and in gas phase (in dashed) as a function of time for the
 reference model (boldface values in Table
 \ref{table_grid}). \textit{Left panel:} the most abundant stable
 species. \textit{Middle panel:} reactive species. \textit{Right
   panel:} less abundant stable species.}
\label{time_Xgrain}
\end{figure*}

This section describes the results obtained using the reference set of
parameters (the boldface values in Table \ref{table_grid} ). Below
 we refer to it as the ``reference model''.  Figure
\ref{time_Xgrain} presents the evolution with time of the abundance of
each species in the gas phase and in the grain mantles.

The evolution of the chemical composition of the mantle depends on the
initial gas phase abundances. In the reference model, O and CO
abundances are approximately five times higher than the H abundance.
Consequently, H atoms cannot hydrogenate all O and CO molecules that
accrete onto the mantles.  Besides, because the CO hydrogenation
reaction has a barrier of 400 K, H atoms react preferentially with O
and its products via barrierless reactions, ending up in water
molecules. In $2 \cdot 10^5$ years, 80 \% of the CO reservoir is
trapped within the mantle bulk, whereas $\sim 60$ \% of O atoms are
contained in water. The remaining O atoms are shared between O$_3$
(ozone), OH and H$_2$O$_2$ (hydrogen peroxyde). Indeed, the high O to
H initial gas phase abundance ratio ($\sim 6$) and the relatively high
temperature (15 K) allow a significant amout of O atoms to diffuse on
the surface and to react with each other before meeting hydrogen atoms.

However, gas phase CO and O abundance ratios relative to H gradually
decrease with time (because they freeze-out onto the mantles), increasing
the possibility of hydrogenation reactions. After $\sim 5 \cdot 10^4$
years, the formation rates of formaldehyde and methanol start to
increase while the formation of ozone stops, corresponding to the time
when H, O, and CO have similar abundances. The percentage of ozone,
formaldehyde, and methanol in the ices therefore strongly depends on the
age of the core. Young cores would show high abundances of ozone and
of hydrogen peroxyde with respect to water, whereas older ones would
show high abundances of formaldehyde and methanol.

Another important result of the model is that a significant amount of
radicals is trapped in the bulk because of the multilayer
treatment. Radicals assumed to be the precursors of COMs such as OH,
HCO or CH$_3$O, reach abundances between $\sim 5 \cdot 10^{-6}$ and
$10^{-9}$. OH is more abundant because of its barrierless formation
reaction, and its formation mostly occurs at shorter times.

\begin{figure}[htp]
\centering
\includegraphics[width=88mm]{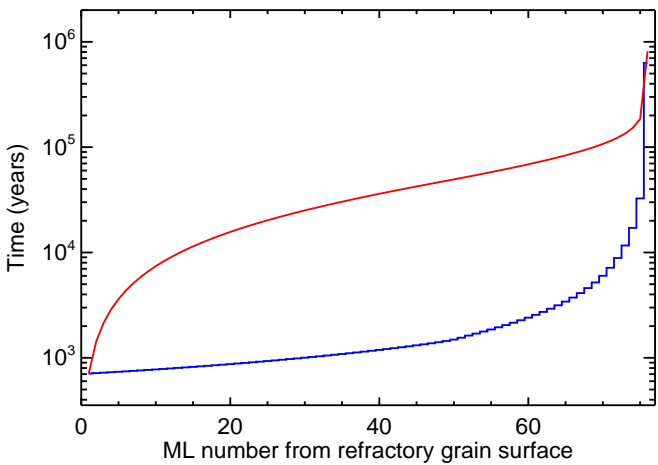}
\caption{Formation time of each monolayer (blue line) and mantle
 formation time versus its thickness expressed in monolayers (red
 line) for the reference model (boldface values in Table
 \ref{table_grid}).}
\label{time_ML}
\end{figure}

The evolution of the chemical composition can also be studied
``spatially". Indeed, our multilayer approach allows us to study the
composition of each monolayer within the grain mantle.  Figure
\ref{time_ML} shows the evolution of the formation time of each layer
and the evolution of the mantle thickness with time. In the reference
model, a grain mantle of 77 layers is formed in $2-3 \cdot 10^5$
yr. Assuming that the layer thickness is equal to the site size (here
4.2 $\AA$), it gives a total mantle thickness of $\sim 0.04~\mu$m,
namely 32 \% of the grain radius. The first 60 layers are created in
less than $10^5$ yr because they have a fast formation rate (each layer
is formed in less than $\sim 2 \cdot 10^3$ yr). The formation time of
layers increases sharply at $\sim 10^5$ yr, because of the drop of the gas
phase abundances.

\begin{figure*}[htp]
\centering
\includegraphics[width=180mm]{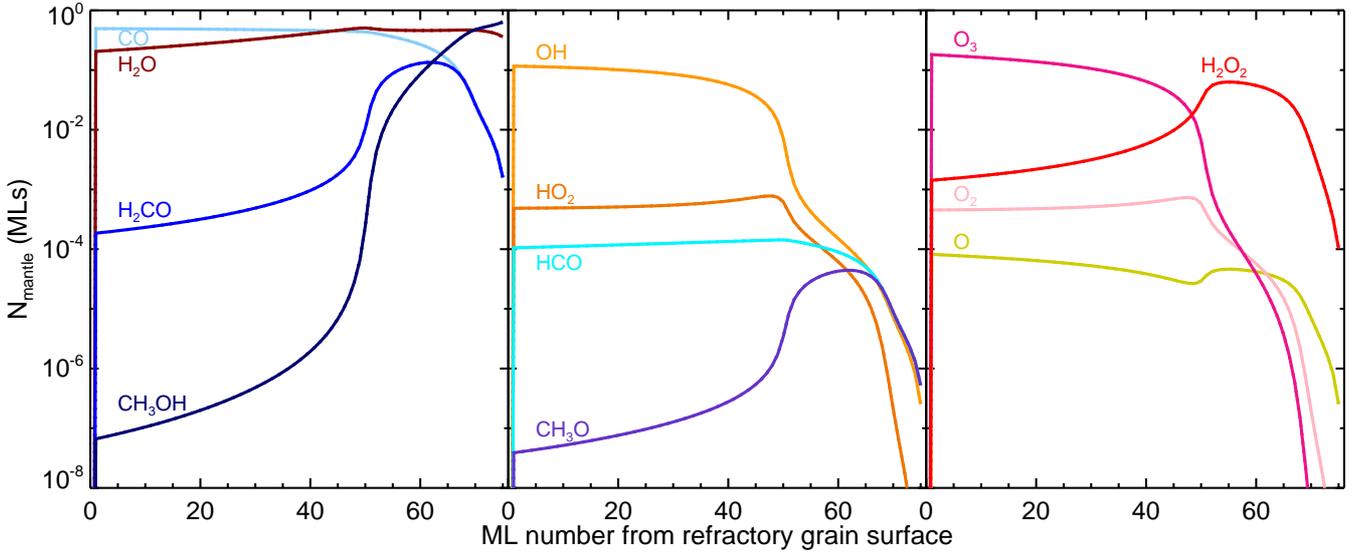}
\caption{Fractional composition of each mantle monolayer for the
 reference model (boldface values in Table \ref{table_grid}).}
\label{ML_compo}
\end{figure*}

The differentiation of the composition between the inner and the outer
layers of the mantle can also be seen in Fig.  \ref{ML_compo}. Indeed,
``intermediate'' molecules such as O$_3$, which is created from
atomic oxygen, are rapidly formed and are therefore abundant in the inner
layers of the mantle (the first 50 layers), whereas formaldehyde
and methanol are mostly formed later, their fractional composition
showing significant abundances only in the 20 outermost layers. The
abundances of the two main species, CO and H$_2$O, remain relatively
constant throughout the mantle up to the last outermost layers, where
they decrease in favour of H$_2$CO and CH$_3$OH.  The composition of
radicals within the mantle follows the same evolution as their
precursor molecules: the HCO abundance stays relatively constant until
the drop in the last layers like CO, while the OH abundance decreases
following the O abundance behaviour.  Finally, the CH$_3$O abundance
increases approaching the surface, as H$_2$CO.

\subsection{Multilayer versus bulk approach}\label{sec:multilayer-vs-bulk}

\begin{figure*}[htp]
\centering
\includegraphics[width=180mm]{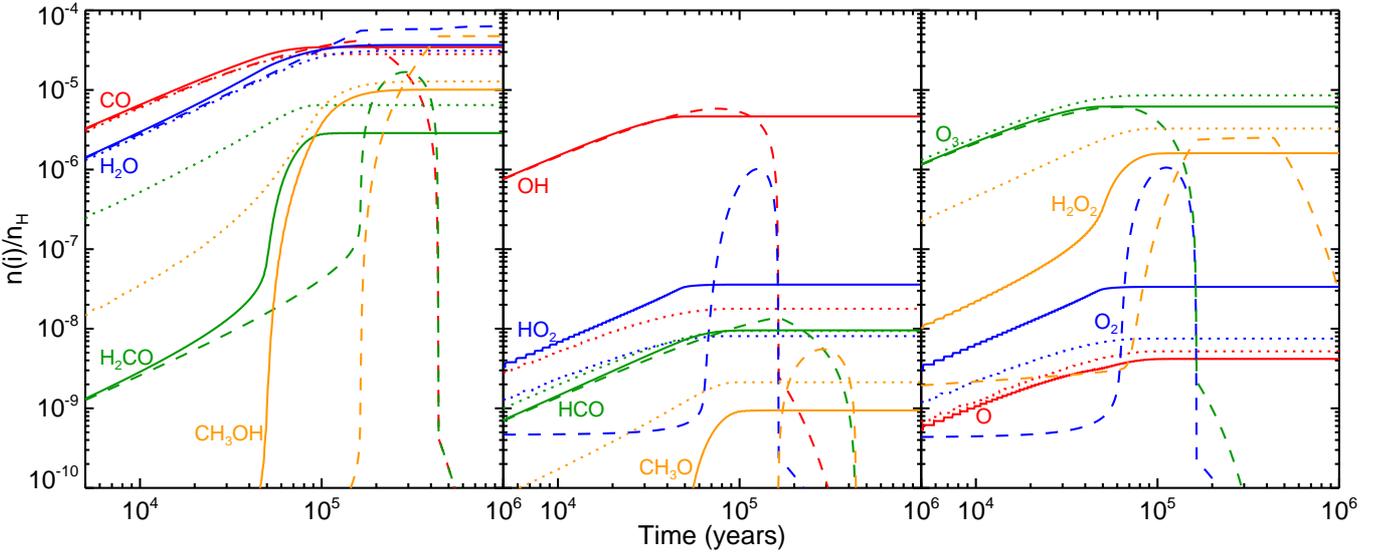}
\caption{Abundance of species in interstellar grain mantles as a
 function of time for the reference model (solid lines), the bulk
 approach model (\S \ref{sec:multilayer-vs-bulk}; dashed lines), and
 the porous grain model (\S \ref{sec:infl-grain-poros}; dotted
 lines). \textit{Left panel:} the most abundant stable
 species. \textit{Middle panel:} reactive species. \textit{Right
   panel:} less abundant stable species.}
\label{time_bulkvslayer}
\end{figure*}

The introduction of the multilayer approach modifies the chemical
behaviour of grain mantles and thus their overall chemical
composition. Figure \ref{time_bulkvslayer} shows the evolution of the
mantle composition with time for the reference model,
adopting a multilayer (described in sec. \ref{sec:multilayer}) or a bulk (old method described in 2.1) approach, respectively.
The figure shows that at short timescales ($\leq 5\times10^4$ yr),
both approaches give similar abundances for the main species 
because the H abundance is initially low compared to the CO and O
abundances (\S \ref{sec:reference-model}).  In contrast, O and its
associated ``intermediate'' molecules (OH, O$_2$, and HO$_2$) are more
efficiently burned in O$_3$, H$_2$O$_2$, and H$_2$O in the bulk compared
 to the multilayer approach.

At later times ($\geq 5\times10^4$ yr), reactive species, which are
trapped by the multilayer approach, continue to react in the bulk
method. Consequently, in the bulk approach, radicals (OH, HCO, HO$_2$,
CH$_3$O) and ``intermediate'' stable species (O, O$_3$, H$_2$O$_2$,
O$_2$) are totally burned in less than $\sim10^6$ yr. In contrast,
the abundances of these species do not evolve after $\sim3 \cdot 10^5$
yr with the multilayer approach. Finally, the predicted methanol and
formaldehyde abundances substantially diverge in the two approaches
at times longer than $\sim 10^5$ yr, namely the presumed ages of the
prestellar cores. The bulk method predicts a final (at $10^6$ yr)
methanol abundance higher by factor 6 compared to the multilayer
approach, and a formaldehyde abundance lower by a factor higher than 10$^4$.

\begin{figure}[htp]
\centering
\includegraphics[width=88mm]{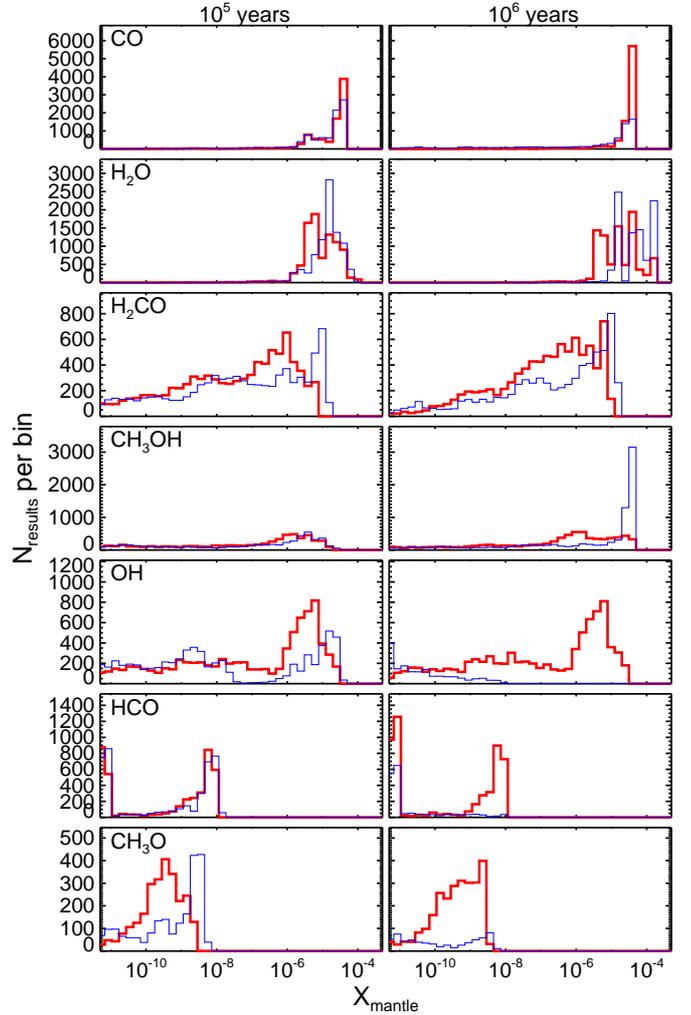}
\caption{Distribution of the predicted mantle abundance $X_{mantle}$ of the key
 species
  at $10^5$ (left panels) and $10^6$ yr (right panels). 
  The thin blue and thick red lines refer to the bulk
 and multilayer approach. The distribution has been
 built by considering all the $\sim$18000 runs of the grid (\S
 \ref{sec:multiparam}).}.
\label{distrib_treat}
\end{figure}

To understand how robust these results are, we considered the whole
grid of models described in \S \ref{sec:multiparam} and built the
distribution of the predicted abundances. Figure \ref{distrib_treat}
shows this distribution for key stable species (CO, H$_2$O, H$_2$CO,
and CH$_3$OH) and the radicals (OH, HCO, and CH$_3$O) for
the bulk and multilayer approaches for two times ($10^5$
and $10^6$ yrs). The plots of Fig. \ref{distrib_treat} tell us that at
short times ($10^5$ yr) the predicted species abundance distributions
are quite similar in the bulk and multilayer approaches, with the
exceptions of H$_2$CO and CH$_3$O, for which the distribution are
slightly different. In contrast, at long times ($10^6$ yr), the
distribution of all radicals are substantially different in the two
approaches, demonstrating that the age is a key parameter.  The plots
also provide another important information. For example, considering
the multilayer approach results, CO and OH have quite peaked
distributions and their predicted abundances depend slightly if at all,
on the assumed values of the parameters: they are robust
predictions. On the contrary, H$_2$CO, CH$_3$OH, HCO and CH$_3$O have
broad abundance distributions implying that their predicted abundances
are very sensitive to the values of the other model parameters.

\begin{normalsize}
\begin{table*}[htp]
\caption{Summary of the effect of each free parameter on the surface chemistry and on the distribution of results (shown in the appendix).}
\begin{tabular}{>{\centering}p{2cm}>{\raggedright}p{5cm}>{\raggedright}p{5cm}>{\raggedright}p{5cm}}
\hline 
\hline
Parameter & Effect on the surface chemistry & Distributions of abundances for stable species & Distributions of abundances 

for radicals\tabularnewline
\hline
Multilayer versus bulk approach & The multilayer approach

- decreases the formation of stable species

- traps reactive species in the mantle & - $X$(CO) $> 10^{-5}$ for 

88\% (ML), 50\% (bulk) cases

- $X$(H$_{2}$O) $> 10^{-5}$ for

60\% (ML), 93\% (bulk) cases 

- $X$(CH$_{3}$OH) $> 10^{-5}$ for

15\% (ML), 55\% (bulk) cases 
& - X(OH) $> 10^{-9}$ for

75\% (ML), 1\% (bulk) cases

- $X$(HCO) $> 10^{-9}$ for

25\% (ML), 2\% (bulk) cases
\tabularnewline
\hline
Porosity  & The increase of porosity

- traps volatile and diffusing particles

- increases the formation rate of stable species & - $X$(H$_{2}$CO) $> 10^{-7}$ for

52\% (smooth), 61\% (porous) cases

- $X$(CH$_{3}$OH) $> 10^{-7}$ for

51\% (smooth), 61\% (porous) cases
& - X(OH) $> 10^{-6}$ for

55\% (smooth), 20\% (porous) cases
\tabularnewline
\hline
Density & The increase of density

- decreases the formation of stable species formed by hydrogenation
reactions & - $X$(H$_{2}$O) $>$ 10$^{-5}$ for

2\% (dense) , 85\% (sparse) cases with $X$(H$_{2}$O) $\rightarrow 2\cdot10^{-4}$

- $X$(CH$_{3}$OH) $> 10^{-5}$ for

0\% (dense) , 35\% (sparse) cases with $X$(CH$_{3}$OH) $\rightarrow 9\cdot10^{-5}$
& - $X$(OH) $> \cdot10^{-6}$ for

0\% (sparse), 75\% (dense) cases 
\tabularnewline
\hline
Temperature  & The increase of temperature

- strongly increases the desorption of H and O

- decreases the rates of hydrogenation reactions

- decreases the abundance of stable species formed by these reactions & - $X$(CH$_{3}$OH) $> 10^{-7}$ for

65\% (cold), 45\% (warm) cases

- $X$(H$_{2}$CO) $> 10^{-7}$ for

60\% (cold), 45\% (warm) cases & - $X$(OH) $> 10^{-6}$ for

55\% (cold), 48\% (warm) cases\tabularnewline
\hline
Grain size 

$a_{d}$ & The increase of the grain size

- increases the mantle thickness

- does not influence the integrated mantle composition & Same evolution of distributions for all stable species & Same evolution of distributions for all radicals\tabularnewline
\hline 
Initial abundance of atomic oxygen & The increase of $X_{ini}$(O):

- increases the formation of water (and other molecules formed from
reactions involving O)

- slightly decreases the formation of molecules formed from CO & - $X$(H$_{2}$O) $> 2 \cdot 10^{-5}$ for

0\% (low-O), 65\% (high-O) cases with $X$(H$_{2}$O) $\rightarrow 2\cdot10^{-4}$

- No influence is seen for CO, H$_{2}$CO and CH$_{3}$OH & - $X$(OH) $>4\cdot10^{-6}$ for

0\%  (low-O), 50\% (high-O) cases with $X$(OH) $\rightarrow 4\cdot10^{-5}$ \tabularnewline
\hline
Diffusion energy & The increase of the diffusion energy

- decreases the diffusion rate of mobile species

- decreases the formation rate of stable species

- slightly increases the survival of radicals & - $X$(H$_{2}$O) $>5\cdot10^{-6}$ for

87\% (fast), 74\% (slow) cases

- $X$(CH$_{3}$OH) $> 10^{-7}$ for

80\% (fast), 30\% (slow) cases

- $X$(H$_{2}$CO) $> 10^{-7}$ for

75\% (fast), 40\% (slow) cases & Same evolution of distributions for all radicals \tabularnewline
\hline
Activation energies  & The increase of activation energies

- strongly decreases the formation rate of H$_{2}$CO and CH$_{3}$OH

- strongly decreases the survival of radicals HCO and CH$_{3}$O

- slightly increases the reaction rates of reactions involving O

- slightly increases the formation of water & - $X$(CO) $> 10^{-5}$ for

35\% (low-Ea), 98\% (high-Ea) cases

- $X$(H$_{2}$CO) $>10^{-7}$ for

83\% (low-Ea), 25\% (high-Ea) cases

- $X$(CH$_{3}$OH) $>10^{-6}$ for

80\% (low-Ea), 5\% (high-Ea) cases & - $X$(HCO) $>10^{-10}$ for

90\% (low-Ea), 0\% (high-Ea) cases

- $X$(CH$_{3}$O) $>10^{-10}$ for

70\% (low-Ea), 0\% (high-Ea) cases\tabularnewline
\hline
Site size  & The increase of the site sizes

- decreases the formation time of layers

- increases the mantle thickness

- does not influence the integrated mantle composition & Same evolution of distributions for all stable species & Same evolution of distributions for all radicals\tabularnewline
\hline
\end{tabular}
\label{summary_grid}
\end{table*}
\end{normalsize}

\subsection{Porous versus non-porous grains}\label{sec:infl-grain-poros}

The influence of the grain porosity (\S \ref{sec:porosity}) is shown
in Figure \ref{time_bulkvslayer} for the reference set of parameters.
The presence of the porosity in the grain has the effect of slightly
speeding up and increasing (by less than a factor 4) the formation of
formaldehyde, methanol, and CH$_3$O, as well as to increase the
destruction of HO$_2$, O$_2$ and O.  The reason is that porous grains
trap H atoms more efficiently, increasing their number on the grains,
but in the end, the H atom numbers is still limited by their
low gas phase abundance (Fig. \ref{time_bulkvslayer}). 
Furthermore, the porosity only slightly modifies the distribution of the results
shown in Fig. A.1.a. Therefore, the porosity only enhances
the formation of hydrogenated molecules by a few factors at most, 
regardless of the values of other parameters.

\subsection{Influence of physical conditions}

\noindent
{\it Density $n_H$ }

The density plays an important role for two reasons. First, increasing
the density reduces the time needed to form each monolayer (because the
accretion rate is proportional to the square of the density). Second, the ratio
between the number of H atoms and the number of heavy particles (CO
and O) that land on the grain is inversely proportional to the density
(see section \ref{sec:gasphase}).  As a consequence, the higher the
density, the lower the abundance of stable species (H$_2$O, H$_2$CO,
CH$_3$OH) created by the hydrogenation reactions, especially at long
times ($\geq10^5$ yr) (Fig. A.1.b). In contrast and for the same
reasons, mantle CO and OH are more abundant for higher density.\\

\noindent
{\it Temperature $T_g = T_d $}

The grain temperature has a moderate influence on the predicted mantle
composition, although the diffusion (and thus the reaction) and the
desorption rates depend exponentially on it. However, the desorption
rate increases much more quickly than the diffusion rate with
temperature (by factors of $10^2 - 10^6$ depending on the $E_d/E_b$
ratio). Consequently, with increasing grain temperature, H atoms have
a higher probability to desorb into the gas phase before encountering
another particle. The final mantle abundance of molecules created by
hydrogenation reactions (H$_2$CO and CH$_3$OH) is therefore lower at any
time at higher temperatures, as shown by Fig. A.1.c. 
The mantle abundance
of radicals, on the other hand, is not affected by the grain
temperature.\\

\noindent
{\it Grain size $a_d$}

While the mantle thickness depends roughly quadratically on the grain
size, the final mantle composition depends on it very little. The
reason is that the particle accretion rate (inversely proportional to
the grain size square) and the formation time of each layer compensate
each other. The result is that the mantle composition does not depend
on the assumed grain size, whether 0.1 or 0.3 $\mu$m.

\subsection{Influence of other key parameters of GRAINOBLE}

\noindent
{\it Initial abundance of atomic oxygen $X$(O)$_{ini}$}

Not surprisingly, the initial gas phase abundance of atomic oxygen is an
important parameter for the final mantle water abundance. The higher
the O abundance, the higher the iced H$_2$O. 
Less evidently, oxygenation is in competition with hydrogenation. The 
initial O abundance also affects the final mantle abundance of 
formaldehyde and, even more, methanol, because the elemental C/O ratio 
decreases between 0.7 and 0.2 when $X($O$)_{ini}$ increases between 
$2\cdot 10^{-5}$  and $2\cdot 10^{-4}$ . At 10$^5$ yr, the predicted
methanol abundance is a factor ten higher when the O abundance is a
factor ten lower.\\

\noindent
{\it Diffusion energy to binding energy ratio $E_d/E_b$} 

The diffusion energy is an important parameter in the predicted mantle
abundance of formaldehyde and methanol.  Because the diffusion rates depend
exponentially on the diffusion energy, the mantle abundance of stable
species, like formaldehyde and methanol, decreases with increasing
$E_d$.  Conversely, the diffusion energy does not affect the mantle
abundance of radicals, because their formation and destruction rates
compensate each other.\\

\noindent
{\it Activation energy $E_a$}

The activation energy of CO and H$_2$CO hydrogenation reactions
strongly influences the formation rates of formaldehyde and methanol
because their reaction rates depend exponentially on $E_a$.  Most of the runs
with a low value of $E_a$ (400 K) predict mantle abundances of
formaldehyde and methanol higher than $10^{-6}$ at $\geq 10^5$ yr.
Conversely, higher values of $E_a$ give uniform distributions of
H$_2$CO and CH$_3$OH abundances, between $10^{-18}$ and
$10^{-6}$. Only $\sim 5$\% of the runs using a activation energy
equal to 2500 K predict formaldehyde and methanol abundances higher
than $10^{-6}$. The final abundances of formaldehyde and methanol,
consequently strongly depend on the other model parameters for high $E_a$
values.  A significant amount of radicals, like HCO and CH$_3$O, can
survive in the mantle only if the activation energies are low.  Runs
with activation energies of 400 K predict radical mantle abundances
between $10^{-10}$ and $10^{-8}$ for both radicals, while runs with
$E_a = 2500$ K predict abundances less than $\sim 10^{-11}$. Not
surprisingly, the activation energy of CO and H$_2$CO hydrogenation
reactions are, therefore, critical parameters for the survival of
radicals in the mantle. \\

\noindent
{\it Site size $d_s$}

The reaction rates and the formation time of each monolayer are both
functions of the site size $d_s$: the former is proportional to
$d_s^{2}$, whereas the latter is proportional to $d_s^{-2}$.
Therefore, the two processes, which are in competition, cancel each
other, and consequently, the total mantle abundance of stable species
and radicals do not sensitively depend on the size of sites.

However, the thickness of the grain mantle strongly depends on the site 
size. Indeed, the maximum number of monolayers is given by the
ratio between the number of particles on the grain $N_{part}$ and the
number of grain sites $N_s$ (proportional to $d_s^{-2}$). Furthermore,
if we assume that the thickness of a layer is equal to the
distance between two sites, the thickness of the mantle is proportional
to $d_s^3$.  Thus, the bigger are the sites, the faster the grains grow.
Consequently, the depletion is more efficiently on grains with big sites,
increasing the abundance of methanol and decreasing the abundance of CO
for example.

\subsection{Concluding remarks}
Based on the previous paragraphs, we can identify three classes of
parameters, depending on their influence on the results:
\begin{itemize}
\item 1) The parameters that have a {\it strong} influence on the
 mantle composition are the approach to model the chemical behaviour
 (multilayer versus bulk), the density, the diffusion energy and
 the activation energy of CO and H$_2$CO hydrogenation
 reactions. Modifying the values of these parameters within their
 chosen range drastically changes the distribution of the predicted
 mantle abundances, both in the shape and in the values.
\item 2) The grain porosity and the temperature as well as the initial
 abundance of atomic oxygen have only a {\it moderate} impact on the
 predicted mantle composition.
\item 3) The grain and the site sizes have a {\it negligible}
 influence on the total mantle composition. However, these parameters
 strongly modify the ice thickness, which could have an impact on
 photolytic processes and for the desorption time of the mantle
 during the warm-up phase in protostar envelopes.
\end{itemize}

\section{Comparisons with previous microscopic models}\label{sec:comp-with-prev}

The Monte-Carlo continuous time random walk (CTRW) method was
introduced by \citet{Chang2005} and \citet{Cuppen2005} for the
hydrogen recombination system, and was then extended to more complex
networks in \cite{Chang2007}, \citet{Cuppen2007},
\citet{Cuppen2009}. The natural output of this model is the multilayer
structure of the mantle.  This section compares our results with the
model of \citet{Cuppen2009} with the twofold aim of validating the
GRAINOBLE code and of highlighting the differences between the two
models.

\subsection{Validation of the GRAINOBLE code}

\citet{Cuppen2009} considered the accretion of H, CO, and H$_2$ on
smooth surfaces as described by \citet{Cuppen2005}, leading to the 
formation of formaldehyde, methanol, and their associated radicals 
HCO and CH$_3$O via the Langmuir-Hinshelwood and the Eley-Rideal mechanisms. 
A major difference between GRAINOBLE and the Cuppen et al. model is
that they consider an $\alpha$-CO ice, whereas we assumed that the
bulk of the mantle is formed by iced water.  To reproduce the
interactions of molecules within an $\alpha$-CO ice, we modified the
set of binding energies in our model accordingly. Second, the CTRW
model takes into account the individual interaction energies between
each particle, while our model only considers the total binding energy
between the adsorbate and the substrate. To simulate their approach,
we assumed for each species $i$ a binding energy equal to fourteen
times the interaction between the species $i$ and a CO molecule
($E_b(i) = 14 E_{i-CO}$), corresponding to the interaction between an
adsorbate and a porous multi-layered CO ice. The binding energy of H
is equal to 450 K, while the binding energy of CO is 880 K.  The
diffusion to desorption energy ratio is assumed to be 0.8, as in
\citet{Cuppen2009} for a layer that is initially smooth.  The
activation energies for CO and H$_2$CO hydrogenation reactions are
those measured by \citet{Fuchs2009}, and slightly depend on the grain
temperature.

\begin{figure}[htp]
\centering
\includegraphics[width=88mm]{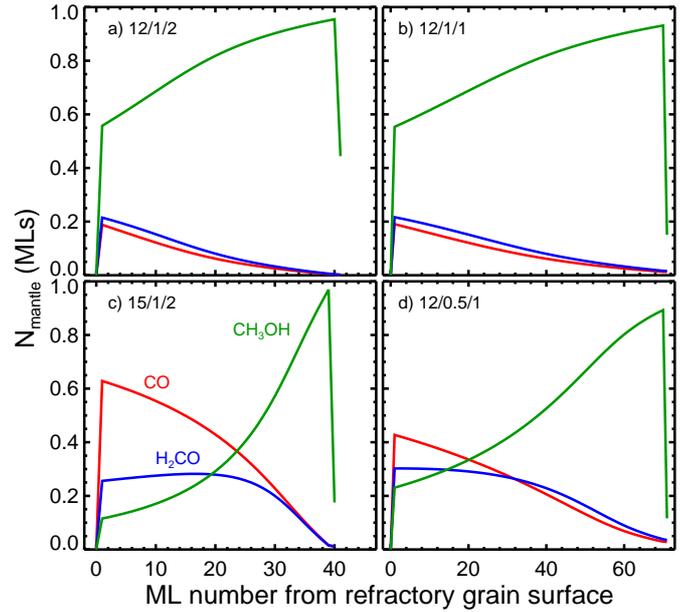}
\caption{Chemical composition of the mantle layers as a function of
 the monolayers at a time of $2 \cdot 10^5$ years, for a 
 density $n_H = 10^5$ cm$^{-3}$, as in \citet{Cuppen2009}: see
 text): CO (red), formaldehyde (blue) and methanol (green). The four
 panels refer to the four cases of Fig.4 of \citet{Cuppen2009} (see
 text): a) grain temperature 12 K, $n(H)/n(CO)$=1 and $n_d = 2\times
 10^{12} n_h$ ; b) as a) but $n_d = 1\times 10^{12} n_h$; c) as a)
 but temperature 15 K; d) as b) but $n(H)/n(CO)$=0.5.}
\label{results_cuppen}
\end{figure}

To validate our code with respect to the model of \citet{Cuppen2009}, we run our model
with a density of $10^5$ cm$^{-3}$ and considered the results at
$2\times10^5$ yr, for the following four cases (as in Part 3.3 of
\citet{Cuppen2009}): a) grain temperature 12 K, $n(H)/n(CO)$=1 and
$n_d = 2\times 10^{12} n_h$ ; b) as a) but $n_d = 1\times 10^{12}
n_h$; c) as a) but temperature 15 K; d) as b) but $n(H)/n(CO)$=0.5.
Figure \ref{results_cuppen} shows the chemical composition of each
monolayer for the four cases, for porous grains ($F_{por} = 0.7$: this
is the value that agrees best for our and Cuppen's
model).  The comparisons between Figure 4 of \cite{Cuppen2009} and our
Figure \ref{results_cuppen} shows that the two codes give very similar
results, which validates our code.

\subsection{Differences between the two approaches}

We highlight a few critical differences between the model of
Cuppen and ours.

Unlike our multilayer approach, the CTRW method allows us to model the
segregation effects. Heavier particles can agglomerate together to
form islands where they can be fixed onto the grains more
efficiently. Our multilayer method is a macroscpic approach where
segration effects are not taken into account. However, the pores can
play the same role as the islands, because particles that enter into
the pores cannot desorb either. This is supported by the fact that our
porous grains case agrees best with Cuppen's model.

An important difference between ours and Cuppen et al.'s model is the
inclusion of the water ice formation. Because Cuppen et al. did not
consider the formation of water, their ices are formed by pure
$\alpha$-CO ice. Consequently, the interaction energies are lower than
those used in our model. Therefore in Cuppen's
model ices can form only at temperatures lower than 16-18 K, whereas in
our model iced CO can still survive until 21-23 K.  In our model,
then, the formation of formaldehyde and methanol can occur in a wider
range of temperatures than in \citet{Cuppen2009}. Consequently, the
abundance of formaldehyde and methanol formed on grain mantles decreases
much more efficiently with the increase of the temperature in Cuppen et 
al.'s model than in ours. For example, using the same physical parameters, the 
CTRW model predicts a decrease of the mantle thickness of 85 \% ($\sim$ 40 to 6 
monolayers), whereas our model does not predict any thickness decreases 
between 15 and 16.5 K.

\section{Comparisons with observations}\label{sec:comp-with-observ}

\subsection{The observations}

In order to compare our model predictions with observations, we used
the data pertaining to the ices only, because the abundance of the sublimated
mantle species does not necessarly reflect the chemical composition of
their precursor ices. First, it can be altered by reactions in the gas
phase. Second, the different sublimation temperatures may introduce
errors when comparing the abundance ratios of different species, because
they may refer to different regions.

Unfortunately, solid CO, H$_2$CO, and CH$_3$OH have been observed
simultaneously only towards a very small sample of high- and
intermediate-mass protostars \citep{Gibb2004,Pontoppidan2004}.  To 
increase the statistics and to include low-mass protostars,
too, we therefore restricted the comparison of our model to the
observed solid CO and CH$_3$OH abundances only. Fortunately, {\it a
 posteriori}, when considering the model predictions, we found that
using the H$_2$CO would not provide (substantially) more constraints.
Table \ref{obs_CO_CH3OH} reports the compilation of observations that
we used for the comparison.
\begin{table}[htp]
\centering
\caption{Mantle abundances of CO and CH$_3$OH with respect 
 to water along with the relative CH$_3$OH/CO abundance ratio, 
 as observed towards low- (LM), intermediate- (IM), and high-mass (HM) protostars. }
\begin{tabular}{c c c c}
\hline
\hline
Source & X(CO) & X(CH$_3$OH) & X(CH$_3$OH) \\
& \% wrt H$_2$O & \% wrt H$_2$O & \% wrt to CO \\
\hline
\textbf{Low-Mass} & & & \\
L1448 IRS 1  &  45.5 \tablefootmark{a-e}	& $<14.9$ \tablefootmark{d}	& $<32.7$ \\ 
L1455 SMM 1 & 5.6 \tablefootmark{a-e}	& $<13.5$ \tablefootmark{d} & $<241.1$ \\
RNO 15         &  13.6 \tablefootmark{a-e}& $<5.0 $ \tablefootmark{d} & $<36.8$ \\  
IRAS 03254    & 13.6 \tablefootmark{a-e}	& $<4.6 $ \tablefootmark{d} & $<33.8$  \\ 
IRAS 03271   &  8.2 \tablefootmark{a-e}	& $<5.6$ \tablefootmark{d} & $<68.3$   \\
B1-a              &  13 \tablefootmark{a-e}	& $<1.9$ \tablefootmark{d} & $<14.6$   \\
B1-c              &  28.6  \tablefootmark{a-e}	& $<7.1$ \tablefootmark{d} & $<24.8$   \\
L1489            &  15.2 \tablefootmark{a-e}	& $4.9$ \tablefootmark{d} & 32.3   \\
DG Tau B       &  20.5 \tablefootmark{a-e}	& $<5.7$ \tablefootmark{d} & $<27.8$  \\
IRAS 12553   &  12.6 \tablefootmark{a-e}		& 	$<3.0$ \tablefootmark{d} & $<23.8$ \\ 
IRAS 13546   &  22.7 \tablefootmark{a-e}		& 	$<3.9$ \tablefootmark{d} & $<17.2$  \\ 
IRAS 15398   &  40.3 \tablefootmark{a-e}		& 	10.3 \tablefootmark{d}	& 25.6   \\
CRBR 2422.8-3423 &  11.2 \tablefootmark{a-e}	& $<9.3$ \tablefootmark{d} & $<83.0$  \\
RNO 91          &  19.0 \tablefootmark{a-e}	& $<5.6$ \tablefootmark{d} & $<29.5$  \\
IRAS 23238    &  4.0 \tablefootmark{a-e}	& $<3.6$ \tablefootmark{d}	& $90.0$  \\
\hline
\textbf{Intermediate-Mass} & & & \\
AFGL989         & 19.0 \tablefootmark{b}	& 1.7 \tablefootmark{b} & 8.9  \\
SMM4              & 30.0 \tablefootmark{c}	& 28.0 \tablefootmark{c}	& 93.3  \\ 
\hline
\textbf{High-Mass} & & & \\
W33A   & 7.4 \tablefootmark{a-e} & 14.7	\tablefootmark{d} & 198.6  \\ 
GL 2136  & 10.2 \tablefootmark{a-e} & 8.5 \tablefootmark{d} & 83.3  \\	
S140  & 16.6 \tablefootmark{a-e} & 	$<3.0$ \tablefootmark{d} & $<18.1$  \\
NGC 7538 IRS 9  & 16.5 \tablefootmark{a-e} & 7.5 \tablefootmark{d} & 45.5  \\	
AFGL2136  & 5.0 \tablefootmark{b} & 5.0	\tablefootmark{b} & 100.0  \\ 
AFGL7009  & 16.0 \tablefootmark{b} & 33.0 \tablefootmark{b} & 206.3  \\ 
\hline
\end{tabular}
\label{obs_CO_CH3OH}
\tablebib{$^{(a)}$ \citet{Pontoppidan2003}; $^{(b)}$ \citet{Gibb2004}; 
 $^{(c)}$ \citet{Pontoppidan2004}; $^{(d)}$ \citet{Boogert2008}; 
 $^{(e)}$ \citet{Pontoppidan2008}.}
\end{table}

\subsection{Constraining the activation and diffusion energies}

\begin{figure}[tbh]
\centering
\includegraphics[width=88mm]{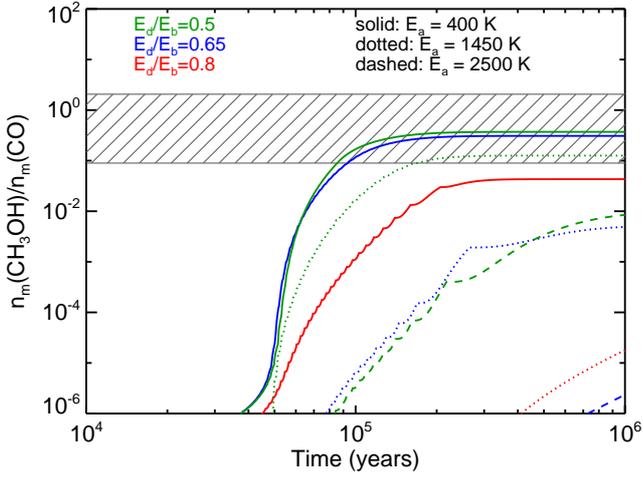}
\caption{Mantle CH$_3$OH/CO abundance ratio versus time. The blue,
 green, and red lines refer to $E_d$/$E_b$=0.65, 0.5, and 0.8,
 respectively. Solid, dotted, and dashed lines refer to $E_a$=400 K,
 1450 K, and 2500 K, respectively. In these computations the density is
 $10^5$ cm$^{-3}$. The box with hatching shows the interval of CH$_3$OH/CO
 abundance ratio observed towards intermediate- and high- mass
 protostars (Table \ref{obs_CO_CH3OH}). Because low-mass protostars
 provide only less stringent upper limits, they are not reported in
 the plot.}
\label{fig:obs-EaEd}
\end{figure}

As a first step, we compared the model predictions with the observations
with the goal to constrain the two microphysics parameters, which have
a high impact on the model predictions (\S 4.6), namely $E_d/E_b$ and $E_a$. Figure
\ref{fig:obs-EaEd} shows the CH$_3$OH/CO abundance ratio as a function
of time obtained assuming reference parameters (and more particularly 
a density $10^5$ cm$^{-3}$) and for different
values of the activation energy $E_a$ and diffusion energy $E_d$.
 The comparison of the model predictions with the values
observed towards intermediate- and high- mass protostars clearly
excludes an activation energy higher than 1450 K and a
diffusion energy higher than 0.65 times the binding energy $E_b$.
More stringently, if $E_a = 1450$ K, $E_d$/$E_b$ has to be equal to
0.5. The analysis of a similar plot for a density of $10^4$ cm$^{-3}$
gives similar constraints: $E_a$=1450 K implies $E_d$/$E_b$=0.5 and
$E_a$=400 K, $E_d$/$E_b\sim 0.65$. The case $10^6$ cm$^{-3}$ does not
provide more constraints, it is indeed excluded by the observations
(see below).

\subsection{Constraining the pre-collapse phase duration and density}

\begin{figure}[tbh]
\centering
\includegraphics[width=88mm]{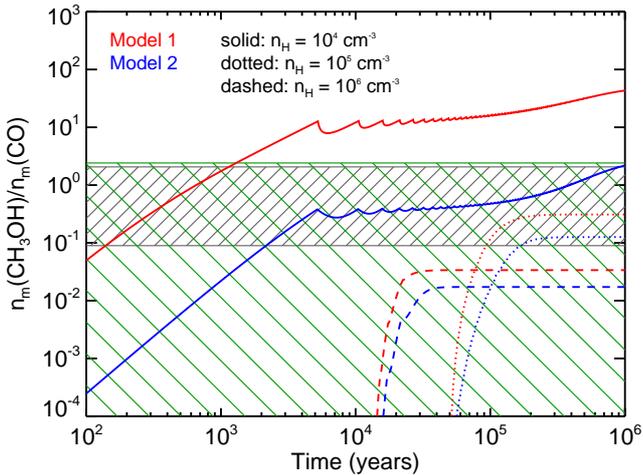}
\caption{Mantle CH$_3$OH/CO abundance ratio versus time. The red lines
 refer to Model 1 ($E_a$=400 K and $E_d$/$E_b$=0.65), the blue lines
 refer to Model 2 ($E_a$=1450 K and $E_d/E_b$=0.5) (see text).
 Solid, dotted and dashed lines refer to densities $10^4$,
 $10^5$ and $10^6$ cm$^{-3}$ respectively. The grey box with hatching shows
 the interval of CH$_3$OH/CO abundance ratio observed towards high-
 and intermediate- mass protostars, while the green dashed box shows
 the values (upper limits) observed in low-mass protostars (Table
 \ref{obs_CO_CH3OH}).}
\label{fig:mod-obs-dens}
\end{figure}

Once we limited the range of possible values of $E_a$ and $E_b$
consistent with the observations, we can attempt to constrain the
duration of the pre-collapse phase and the density. To this end,
Figure \ref{fig:mod-obs-dens} shows the curves of the CH$_3$OH/CO
abundance ratio versus time for three different densities ($10^4$,
10$^5$ and $10^6$ cm$^{-3}$) and computed for the two following sets
of $E_a$ and $E_d$ values: 1) $E_a$=400 K and $E_d/E_b$=0.65 (and 
the reference parameters) ; 2) $E_a$=1450 K and $E_d/E_b$=0.5 (and 
the reference parameters).
The comparison between the predicted and observed CH$_3$OH/CO
abundance ratio in intermediate- and high-mass protostars suggests
 that the bulk of methanol has been formed
when the pre-collapse condensation had densities between $10^4$ and
$10^5$ cm$^{-3}$. Higher densities would produce an insufficiently CH$_3$OH/CO
abundance ratio compared to the values observed in intermediate-
and high- mass protostars. If the densities are around $10^5$
cm$^{-3}$, the duration of the pre-collapse phase must have lasted
at least $10^5$ yr. Lower densities would allow shorter pre-collapse
duration times.  But densities higher than $10^5$ cm$^{-3}$ are
not ruled out for the condensations that produce the low-mass
protostars. In that case, the duration of the pre-collapse phase
lasted more than $5\times10^4$ yr.

One important result is that the CH$_3$OH/CO abundance ratio does not
provide an estimate of the duration of the pre-collapse phase, but
only a lower limit to it. It neither provides an estimate of the
density at the time of the collapse but only a range of possible
densities for any given observed CH$_3$OH/CO abundance ratio. Note
that the CH$_3$OH/H$_2$CO abundance ratio does not allow us to arrive at better
constraints, because the theoretical curves are very similar to those of
Fig. \ref{fig:mod-obs-dens}.

\subsection{Chemical differentiation within grain mantles}

As briefly mentioned in \S 2, IR observations suggest a chemical
differentiation of the grain mantles with a polar matrix formed
already at $A_V \sim 3$ mag \citep{Whittet1988}, and a non-polar CO
matrix formed at higher densities \citep{Tielens1991}. In between, CO
can be mixed either with water, methanol, or both at the same time
\citep{Bisschop2007}, as well as mixed with O$_2$, O$_3$, N$_2$, or
CO$_2$ \citep{Pontoppidan2006}. Actually, this observational fact is
one of the motivations of our work. As discussed in \S
\ref{sec:results} and, for example, shown in Fig. \ref{ML_compo}, our
code predicts inhomogeneous grain mantle composition. For the
reference model, for example, CO is mixed with water in the inner
layers while a mixture between CO and methanol is predicted in the
outer layers.  Given the lack of a dataset to compare our results with, it is
difficult to carry out quantitative comparisons between observations
and predictions.  However, it is worth noticing that the exact grain
mantle composition and stratification depends on the overall evolution
and structure of the pre-collapse condensation, because the chemical
differentiation depends on the density, the temperature and time.

\section{Conclusions and perspectives}\label{sec:concl-persp}

We presented a new model, GRAINOBLE, which computes the chemical
composition of interstellar grain mantles during the cold and dense
phase of the pre-collapse. The model presents two main differences
from most other published codes: 1) it assumes that only the outermost
mantle layer is chemically active, while the mantle bulk is not; 2) it
considers porous grains.

We run GRAINOBLE for a large set of parameters, which are either
unconstrained (density, temperature and grain size of the prestellar
condensation) or poorly known (gas atomic oxygen abundance,
diffusion energy, hydrogenation reactions activation energy, grain
site size) . We obtained a grid of about 18000 models and built
distribution plots of the predicted mantle abundances. This allows us
to study the influence of each of these parameters on the predicted
abundances and, consequently, the robustness of the predictions.

Finally, we compared the GRAINOBLE predictions with those obtained by
the microscopic CTRW model of \citet{Cuppen2009}, obtaining a fair
agreement between the two models. The comparison clearly validates our
treatment and our code (\S \ref{sec:comp-with-prev}).
The most important results of the GRAINOBLE model are the following \\
1) The multilayer treatment shows a differentiation of the species
within the mantle, with the innermost layers being rich in CO, the
intermediate layers in formaldehyde, and the outermost layers in
methanol, reflecting the different formation time of each species in
the mantle.  This differentiation will likely lead to a
differentiation in the deuteration of formaldehyde and methanol on the
ices. Because the deuteration increases with the CO depletion in the gas
phase \citep{Roberts2003, Ceccarelli2005}, methanol will likely be
more deuterated than formaldehyde, as indeed observed in
\citet{Parise2006}. A forthcoming paper will focus on the detailed
modeling of the deuteration process. \\
2) The multilayer treatment predicts a relatively high abundance of
radicals trapped in the mantle. {For example, HCO and CH$_3$O show abundances
of $10^{-9}-10^{-7}$ with respect to H nuclei. As suggested by Garrod \& Herbst
(2006), these radicals can react to form COMs when the grain
temperature increases during the formation of the protostar. At this
stage, we cannot say whether the predicted radical abundances are
sufficient to explain the present observations (e.g. Fig. 1), and,
unfortunately, comparisons with previous models are not possible
because of the lack of specific information.
We will explore this aspect, the formation of COMs, in a forthcoming paper. \\
3) The presence of porosity in the grains only moderately influences
the mantle chemical composition, causing an enhanced abundance of
formaldehyde and methanol by less than a factor four.\\
4) The chemical composition of grain mantles strongly depends on the
physical conditions of the prestellar condensation, particularly on
its density as well as its age.  Therefore, it will be important to
model the evolution of the condensation for realistic predictions of
the mantle composition. Conversely, the observed mantle composition
can provide valuable information on the past history of protostars,
and will be the focus of a forthcoming paper. Comparisons of the
present predictions with the ices observations (specifically the
CH$_3$OH/CO abundance ratio) suggest that intermediate- and high- mass
protostars evolved from condensations less dense than about $10^5$
cm$^{-3}$, while no stringent constraints can be put on the prestellar
condensations of low-mass protostars.\\
5) The predicted mantle composition critically depends on the values
of the diffusion energy and activation energy of the hydrogenation
reactions.  Comparing our model predictions with observations of ices
allows to constrain the values of these two important parameters: the
diffusion to binding energy ratio has to be around 0.5--0.65, and the
activation energy has to be less than 1450 K.\\
6) Other parameters of the GRAINOBLE model, like the gas atomic oxygen
abundance and the site size, do not substantially influence the
predicted mantle abundances.

In summary, GRAINOBLE is a versatile and fast code to model the grain
surface chemistry. Its main improvement with respect to previous
similar models is the treatment of the layer-by-layer chemistry, which
provides a more realistic stratified mantle composition. It predicts
that the mantles are indeed stratified and that radicals are trapped
inside the mantles. Because it is not expensive from the point of view
of computing time, modeling of complex and more realistic cases, like
the evolution of a prestellar core, are now feasible.

\begin{acknowledgements}

The authors would like to thank
O. Biham, P. Caselli, T. Hasegawa, and V. Wakelam for useful exchanges on grain chemistry modelling,
P. Peters and L. Wiesenfeld for discussions about the physical and chemical processes on grain surfaces,
and X. Tielens for helpful discussions on the abundances in the gas phase.

This work has been supported by l\textquoteright Agence Nationale pour la Recherche (ANR), France (project FORCOMS, contracts ANR-08-BLAN-022).

  \end{acknowledgements}
  
\bibliographystyle{aa}

\bibliography{bib}

 \newpage
\appendix

\section{Distributions of abundances}

As introduced in section 3.8, we have built a model grid in which eight input parameters vary. To study the results of this grid, we computed the distributions of selected species abundances on grain mantles $X_m(i)$, namely the number of results giving $X_m(i)$ in each abundance interval. The following figures show the distributions of abundances for each input parameter and for two extremal values, using the multilayer approach only. 
 
\begin{figure}[htp]
\centering
\includegraphics[width=65mm]{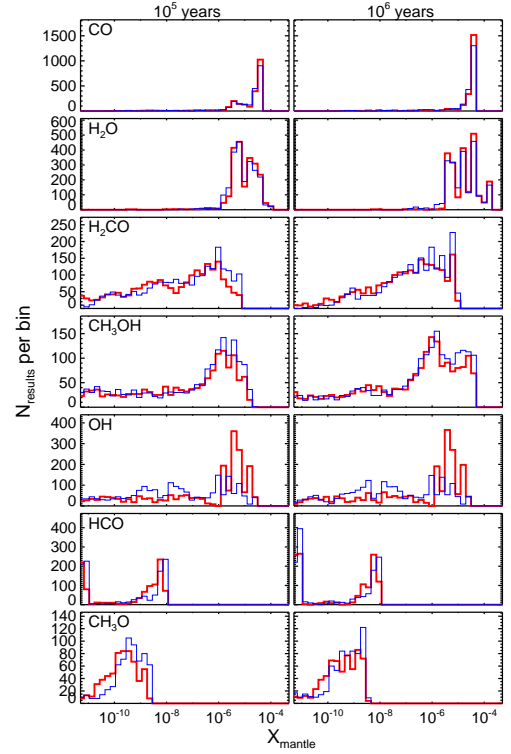}
\caption{Distribution of abundances on grain mantles at $t = 10^5$ and $10^6$ years for two porosity values: $F_{por}$ = 0 (thick red) and 0.9 (narrow blue).}
\label{distrib_Sin}
\end{figure}

\begin{figure}[htp]
\centering
\includegraphics[width=65mm]{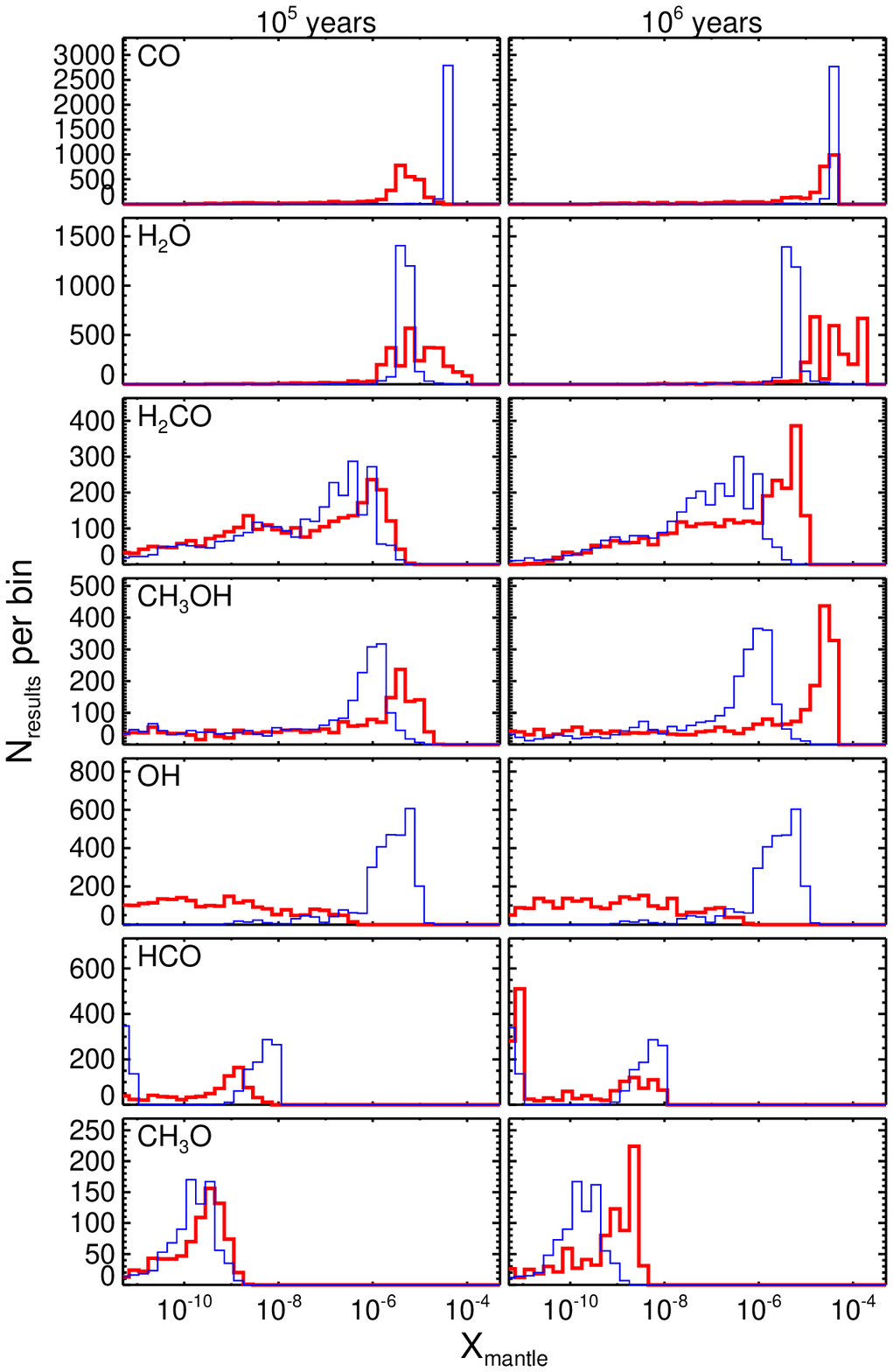}
\caption{Distribution of abundances on grain mantles at $t = 10^5$ and $10^6$ years for two densities: $n_H = 10^4$ (thick red) and $10^6$ cm$^{-3}$ (narrow blue). }
\label{distrib_nH}
\end{figure}

\begin{figure}[htp]
\centering
\includegraphics[width=65mm]{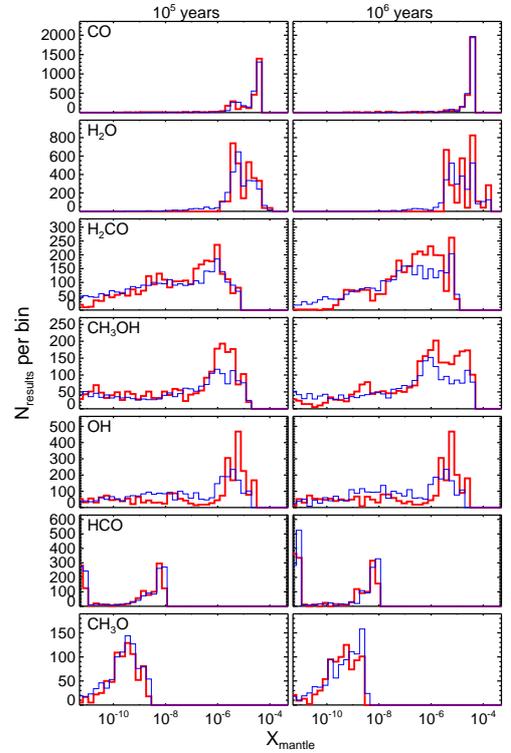}
\caption{Distribution of abundances on grain mantles at $t = 10^5$ and $10^6$ years for two  temperatures: $T = 10$ (thick red) and 20 K (narrow blue).}
\label{distrib_Tg}
\end{figure}

\begin{figure}[htp]
\centering
\includegraphics[width=65mm]{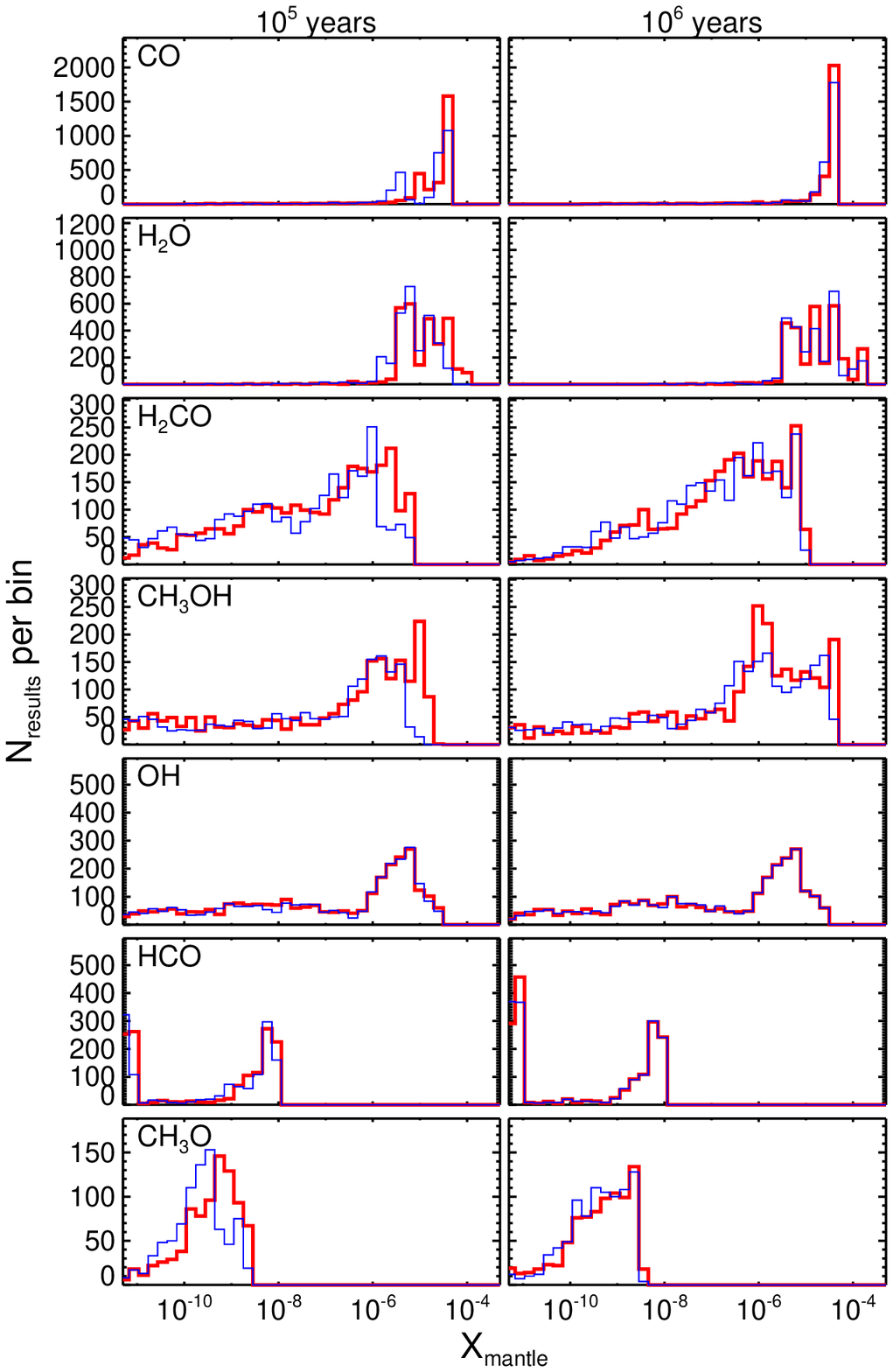}
\caption{Distribution of abundances on grain mantles at $t = 10^5$ and $10^6$ years for two  grain sizes: $a_d = 0.1$ (thick red) and 0.3 $\mu$m (narrow blue). }
\label{distrib_agr}
\end{figure}

\begin{figure}[htp]
\centering
\includegraphics[width=65mm]{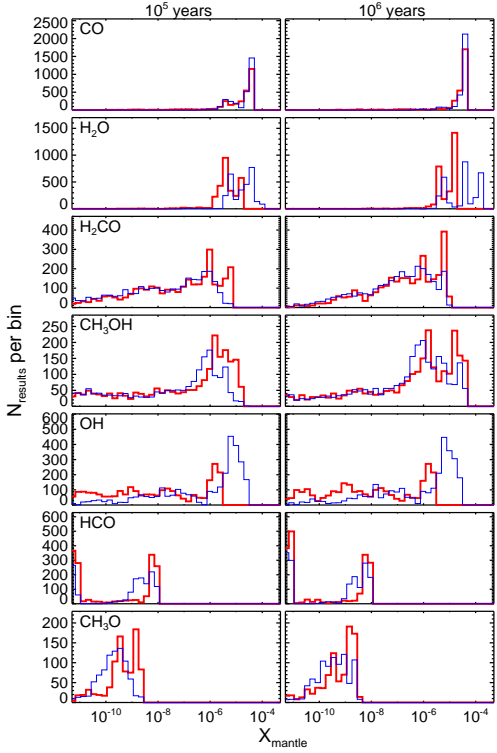}
\caption{Distribution of abundances on grain mantles at $t = 10^5$ and $10^6$ years for two initial oxygen abundances: $X_{ini}$(O) = $2\cdot10^{-5}$ (thick red) and $2\cdot10^{-4}$ (narrow blue).}
\label{distrib_Oini}
\end{figure}

\begin{figure}[htp]
\centering
\includegraphics[width=65mm]{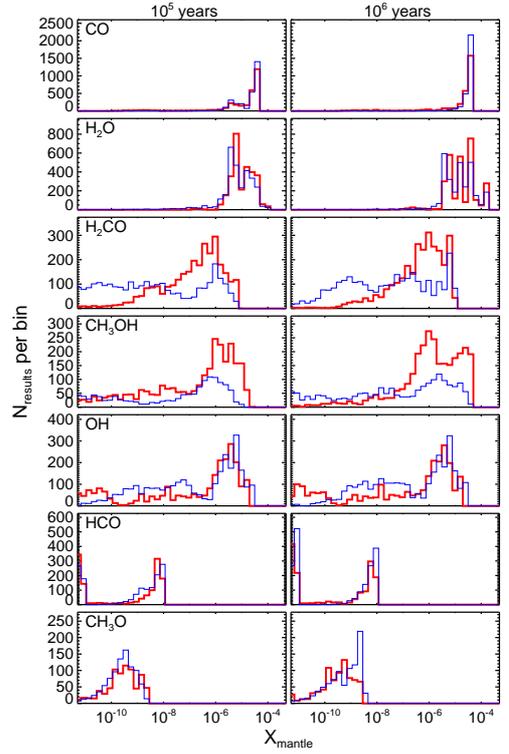}
\caption{Distribution of abundances on grain mantles at $t = 10^5$ and $10^6$ years for two  energy ratios: $E_d/E_b = 0.5$ (thick red) and $0.8$ (narrow blue).}
\label{distrib_enratio}
\end{figure}

\begin{figure}[htp]
\centering
\includegraphics[width=65mm]{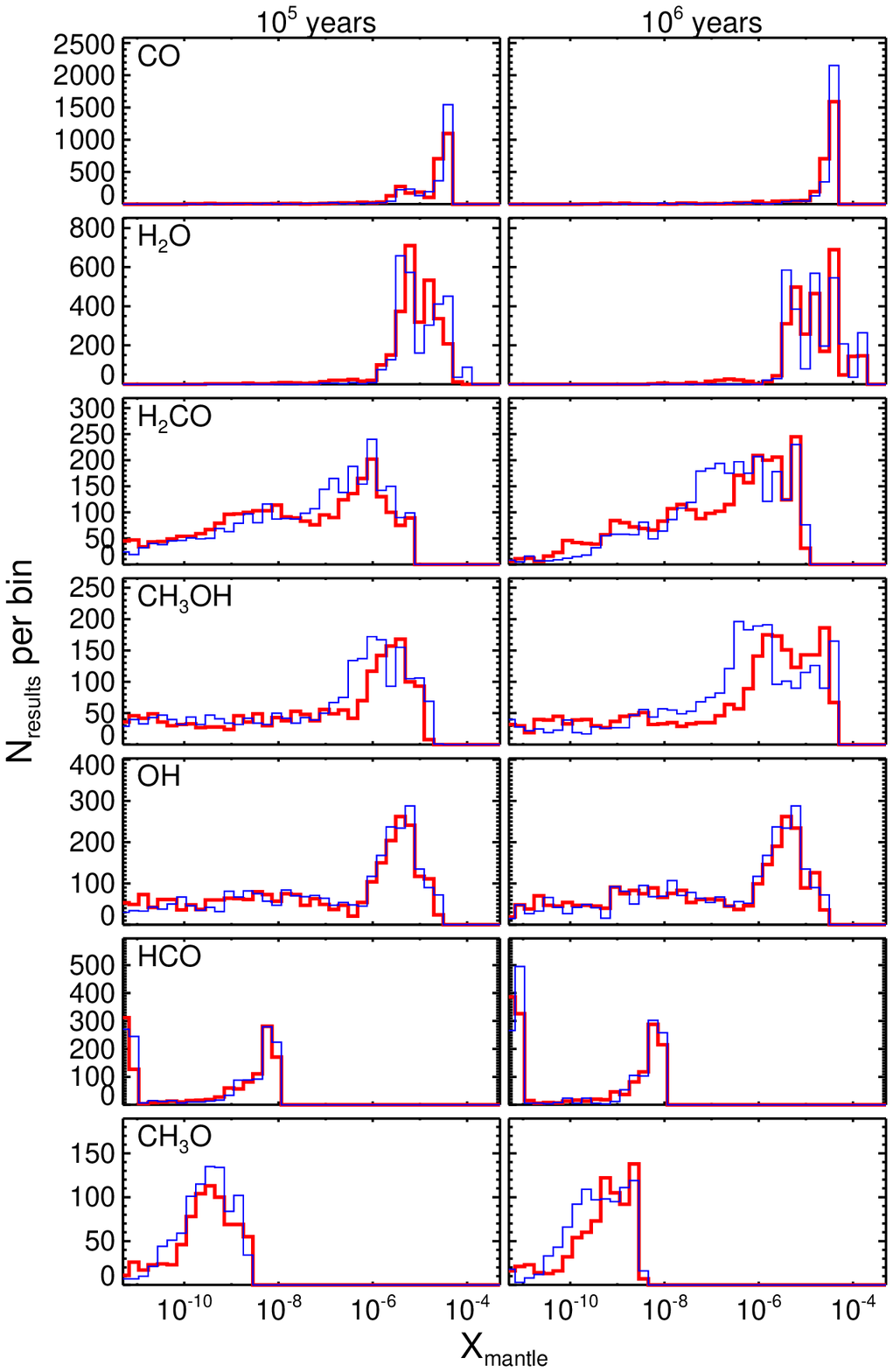}
\caption{Distribution of abundances on grain mantles at $t = 10^5$ and $10^6$ years for two site sizes: $d_s = 1.4$ (thick red) and 7 $\AA$ (narrow blue). }
\label{distrib_ds}
\end{figure}

\begin{figure}[htp]
\centering
\includegraphics[width=65mm]{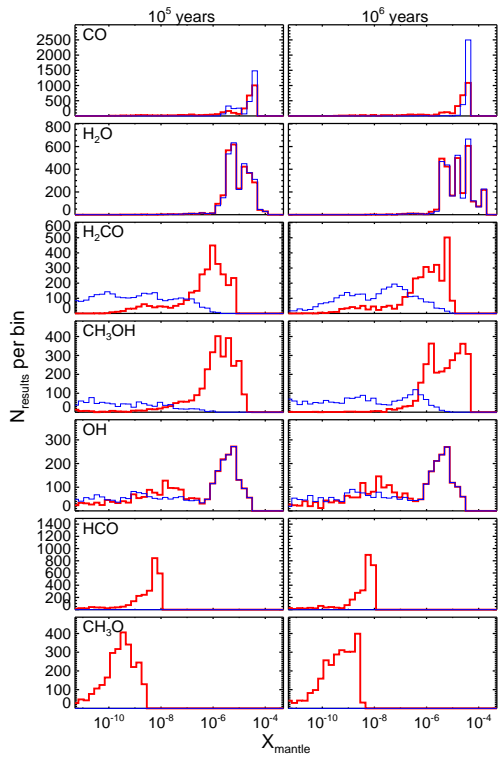}
\caption{Distribution of abundances on grain mantles at $t = 10^5$ and $10^6$ years for two  CO and H$_2$CO hydrogenation reactions: $E_a = 400$ (thick red) and 2500 K (narrow blue).}
\label{distrib_Ea}
\end{figure}

\section{List of symbols used in this work}

\begin{table*}[htp]
\centering
\caption{List of symbols and their explanation.}
\begin{tabular}{l l}
\hline
\hline
Symbol &  Parameter \\
\hline
$a_d$ & diameter of interstellar grains (in $\mu$m) \\
$d_s$ & size of each site (in $\AA$ )\\
$E_d$ & barrier energy against diffusion of a considered species (in K) \\
$E_b$ & barrier energy against desorption of a considered species (in K) \\
$E_a$ & activation energy of the CO and H$_2$CO hydrogenation reactions (in K)  \\
$F_{por}$ & fraction of the grain surface occupied by the pores \\
$F_{np}$ & fraction of the grain surface occupied by the non-porous surface \\
$F_{ed}$ & fraction of the grain surface occupied by the edge porous sites \\
$r_d$ & radius of interstellar grains (in $\mu$m) \\
$n_d$ & density of interstellar grains (in cm$^{-3}$) \\
$N_{pore}$ & number of sites occupied by each pore \\
$N_s$ & number of sites of the considered layer \\
$n_H$ & total density of H nuclei (in cm$^{-3}$) \\
$n_g$ & gas phase density of a considered species (in cm$^{-3}$) \\
$P_{np}$ & surface population of a considered species in the non-porous surface (in monolayers) \\
$P_{por}$ & surface population of a considered species in the pores (in monolayers) \\
$R_{acc}$ & accretion rate of a considered species (in s$^{-1}$) \\
$R_{hop}$ & hopping rate from site to site of a considered species (in s$^{-1}$) \\
$R_{diff}$ & diffusion rate of a considered species (in s$^{-1}$) \\
$R_{r,np}$ & reaction rate of a considered reaction in the non-porous surface (in s$^{-1}$) \\
$R_{r,por}$ & reaction rate of a considered reaction in the pores (in s$^{-1}$) \\
$R_{ev}$ & evaporation rate of a considered species (in s$^{-1}$) \\
$S$ & sticking coefficient of a considered species \\
$\sigma_d$ & cross section of interstellar grains (in cm$^2$) \\
$T_g$ & gas temperature (in K) \\
$T_d$ & dust temperature (in K) \\
$v$ & thermal velocity of a considered species in the gas phase (in cm s$^{-1}$) \\
$X_d$ & abundance of interstellar grains relative to H nuclei \\
$X_{g}$ & abundance relative to H nuclei of a considered species in the gas phase \\
$X_{m}$ & abundance relative to H nuclei of a considered species on grain mantles \\
$\zeta$ & cosmic rays ionization rate (in s$^{-1}$) \\
\hline
\end{tabular}
\label{list_symbols}
\end{table*}

\end{document}